\documentclass[traditabstract]{aa}
\usepackage{natbib}
\bibpunct{(}{)}{;}{a}{}{,}
\usepackage{graphicx}
\usepackage{eufrak}
\usepackage{amsmath}
\usepackage{txfonts}

\begin{document}

\title{Dust-depletion sequences in damped Ly-$\alpha$ absorbers II.}
\subtitle{The composition of cosmic dust, from low-metallicity systems to the Galaxy}
\titlerunning{Dust-depletion sequences in damped Ly-$\alpha$ absorbers II}

\author{Lars Mattsson\thanks{\email{lars.mattsson@nordita.org}}\inst{1}, Annalisa De Cia\inst{2,3}, Anja C. Andersen\inst{4,5} \& Patrick Petitjean$^6$}
\authorrunning{Mattsson et al.}
\institute{Nordita, KTH Royal Institute of Technology and Stockholm University, Roslagstullsbacken 23, SE-106 91, Stockholm, Sweden
\and European Southern Observatory, Karl-Schwarzschild Str. 2, 85748 Garching bei M\"unchen, Germany
\and Department of Particle Physics and Astrophysics, Faculty of Physics, Weizmann Institute of Science, Rehovot 76100, Israel
\and Astrophysics and Planetary Science, University of Copenhagen, \O ster Voldgade 5-7, DK-1350, Copenhagen \O, Denmark
\and Dark Cosmology Centre, Niels Bohr Institute, University of Copenhagen, Juliane Maries Vej 30, DK-2100, Copenhagen \O, Denmark
\and Institut d'Astrophysique de Paris, CNRS and UPMC Paris 6, UMR7095, 98bis Boulevard Arago, F-75014, Paris, France}

\date{Received date; accepted date}

\abstract{
Metals in the interstellar medium (ISM) of essentially all types of galaxies are observed to be depleted compared to the expected values. The depletion is most likely due to dust condensation in, for example, cold molecular clouds and various circumstellar and interstellar environments. The relative observed metal abundances should thereby reflect the composition of the ISM dust components. We aim at identifying the most dominant dust species or types, including silicate and iron oxide grains present in the ISM, using recent observations of dust depletion of galaxies at various evolutionary stages. We use the observed elemental abundances in dust of several metals (O, S, Si, Mg, and Fe) in different environments, considering systems with different metallicities and dust content, namely damped Lyman-$\alpha$ absorbers (DLAs) towards quasars and the Galaxy. We  derive a possible dust composition by computationally finding the statistically expected elemental abundances in dust assuming a set of key dust species with the iron content as a free parameter. Carbonaceous dust is not considered in the present study. Metallic iron (likely in the form of inclusions in silicate grains) and iron oxides are important components of the mass composition of carbon-free dust. The latter make up a significant mass fraction ($\sim 1/4$ in some cases) of the oxygen-bearing dust and there are good reasons to believe that metallic iron constitutes a similar mass fraction of dust. W\"ustite (FeO) could be a simple explanation for the depletion of iron and oxygen because it is easily formed. There appears to be no silicate species clearly dominating the silicate mass, but rather a mix of iron-poor as well as iron-rich olivine and pyroxene. To what extent sulphur depletion is due to sulfides remains unclear. In general, there seems to be little evolution of the dust composition (not considering carbonaceous dust) from low-metallicity systems to the Galaxy.}


\maketitle

\section{Introduction} 
Cosmic dust is an important component of the interstellar medium (ISM) and its study is relevant for planet and star formation, supernova remnants and molecular clouds, as well as cosmology. The composition of dust grains in different environments is an essential part of the puzzle. Different techniques are adopted to infer the properties of dust, including the analysis of dust emission at long wavelengths \citep[see][and references therein]{Draine03} and optical extinction curves and features \citep{Pei92,Cardelli89,Fitzpatrick90}. 

Conventional wisdom had for a long time been that interstellar dust was mostly composed of iron-rich silicates and carbonaceous (graphitic) grains \citep[see, e.g.,][]{Draine84,Draine03,Draine07,Pei92,Weingartner01}. The simple silicate-graphite model (SGM) has remained as the consensus on the origins of the 9.7~$\mu$m and 18~$\mu$m absorption features as well as the ``2175~\AA\ bump'', but more recent results have been shown to be incompatible with the SGM. \citet{Kimura03,Sterken15} showed that the grain-size distribution is likely flatter and more skewed towards large grains than the anticipated distribution in the SGM. Furthermore, space missions (primarily \textit{Cassini} and \textit{Startdust}) found no or little evidence of carbonaceous dust, which raises doubts about graphitic material as the carrier of the 2175~\AA\ extinction feature \citep{Altobelli16,Westphal14}. In addition, the silicates collected in space missions (\textit{Startdust} in particular) tend to be crystalline (forsterite) rather than the amorphous silicate usually assumed in the SGM. New and more sophisticated models have been suggested; for example, an interstellar dust model which includes hydrogenated amorphous carbons as an explanation for the 2175~\AA\ feature and metallic-iron inclusions in silicates to explain the strong depletion of iron, among other things \citep{Jones13,Jones17}.

Another common approach to infer the dust composition is to observe the amount of different metals in the gas-phase ISM. The observations of abundances (with respect to hydrogen) and relative abundances (with respect to other metals), obtained mainly through absorption-line spectroscopy, quantify the amount of metals in the gas-phase of the ISM, Galactic or extragalactic. Some of these metals, and in particular the most refractory elements, are missing from the gas-phase, and are instead incorporated into dust grains. This process is called dust depletion \citep[e.g.,][]{Field74,Cardelli93,Phillips82,Phillips84,Hobbs93,Welty95,Savage96a,Savage96b,DeCia16}. For example, elements like iron and chromium are heavily depleted into dust grains, while oxygen and zinc are little affected by dust depletion.

There are essentially two ways to obtain dust depletion: either metals are depleted where they are formed, for example in supernovae and AGB stars, or depletion is a consequence of dust condensation (chemical grain growth by accretion of molecules) in the ISM \citep{Mattsson16}. Apart from observed depletions, direct evidence for the latter comes from infrared emission from large dust grains in cold molecular clouds, which are thought to be the ``dust factories'' of the ISM \citep[see, e.g.,][]{Steinacker10, Hirashita14}. Indirect support for the interstellar-growth hypothesis comes from destruction of grains in the ISM, highlighting the need for a replenishment mechanism \citep{McKee89,Draine79}. Recently, \citet{Dwek16} showed that interstellar condensation is needed also because there is no other way to explain the high levels of iron depletion in the ISM. In addition, this suggests a much higher iron content in dust than what can be allowed in the SGM. Yet another piece of indirect evidence for this type of grain growth comes from the fact that late-type galaxies seem to have steeper dust-to-gas gradients than metallicity gradients along the radial extension of their disks \citep{Mattsson12a,Mattsson12b,Mattsson14b}.

The study of dust depletion in the Galaxy was expanded and innovated by \citet{Jenkins09}. This work revealed that abundances of several metals correlate with each other, and the author derived continuous sequences of dust depletion, from systems with a small amount of dust, to the most dusty clouds in a Galactic sample. Furthermore, \citet{Jenkins14} used this information to infer ``consumption rates'' in which each element is assembled in dust grains, useful to study dust composition. 

 \begin{table*}
\centering
\caption{Elemental abundances in dust,  $\epsilon_X = \log ({\rm X/H})_{\rm dust}$, for different levels of [Zn/Fe].}
\begin{tabular}{ l | r r r r r r r r r }
\hline \hline
\rule[-0.2cm]{0mm}{0.8cm}
[Zn/Fe]  &   $\epsilon_{\rm Zn}$  &   $\epsilon_{\rm O}$ &  $\epsilon_{\rm P}$  &  $\epsilon_{\rm S}$  &  $\epsilon_{\rm Si}$  &  $\epsilon_{\rm Mg}$  &$\epsilon_{\rm Mn}$  &  $\epsilon_{\rm Cr}$ &  $\epsilon_{\rm Fe}$  \\
\hline

    0.00
 & $-10.69$ 
 & $ -6.50$ 
 & $ -9.57$ 
 & $ -8.06$ 
 & $ -7.19$ 
 & $ -6.81$ 
 & $ -9.28$ 
 & $ -9.60$ 
 & $ -6.96$ 
\\
    0.40
 & $ -9.16$ 
 & $ -5.46$ 
 & $ -8.26$ 
 & $ -6.53$ 
 & $ -5.72$ 
 & $ -5.81$ 
 & $ -7.75$ 
 & $ -7.68$ 
 & $ -5.67$ 
\\
    0.80
 & $ -8.43$ 
 & $ -4.69$ 
 & $ -7.71$ 
 & $ -5.87$ 
 & $ -5.08$ 
 & $ -5.22$ 
 & $ -7.15$ 
 & $ -6.92$ 
 & $ -5.08$ 
\\
    1.20
 & $ -7.97$ 
 & $ -4.02$ 
 & $ -7.42$ 
 & $ -5.52$ 
 & $ -4.73$ 
 & $ -4.77$ 
 & $ -6.85$ 
 & $ -6.55$ 
 & $ -4.72$ 
\\
    1.60
 & $ -7.52$ 
 & $ -3.34$ 
 & $ -7.13$ 
 & $ -5.16$ 
 & $ -4.37$ 
 & $ -4.33$ 
 & $ -6.55$ 
 & $ -6.18$ 
 & $ -4.36$ 
\\

\hline
$\sigma_{\epsilon_X}$ & 0.30 &     0.52 &     0.42 &     0.33 &     0.36  &     0.49  &     0.40  &   0.32  &  0.34 \\
\hline \hline
 \end{tabular}
\label{tabdata}
\end{table*}

  \begin{table*}
  \begin{center}
  \caption{\label{parameters} Compositions and $x$-parameter ranges used for the modeling.}
  \begin{tabular}{lllllllllllll}
  \hline
  \hline
  Species   & Chemi. form.                                    & A1 & B1 & C1 & D1 & & A2 & B2 & C2 & D2 & Stoich. param.\\
  \hline
  Olivine   & Mg$_{2x_{\rm ol}}$Fe$_{2(1-x_{\rm ol})}$SiO$_4$ & $\surd$ & $\surd$ & $\surd$ & $\surd$ & & $\surd$ & $\surd$ & $\surd$ & $\surd$ & $x_{\rm ol} \in [0, 1.0]$\\
  Pyroxene  & Mg$_{x_{\rm py}}$Fe$_{1-x_{\rm py}}$SiO$_3$     & $\surd$ & $\surd$ & $\surd$ & $\surd$ & & $\surd$ & $\surd$ & $\surd$ & $\surd$ & $x_{\rm py} \in [0, 1.0]$ \\
  Forsterite   & Mg$_{2}$SiO$_4$ & - & - & - & - & & $\surd$ & $\surd$ & $\surd$ & $\surd$ & $x_{\rm ol} = 1.0$\\
  Enstatite  & MgSiO$_3$     & - & - & - & - & & $\surd$ & $\surd$ & $\surd$ & $\surd$ & $x_{\rm py} = 1.0$\\
  Iron      & Fe                                              & $\surd$ & $\surd$ & $\surd$ & $\surd$ & & $\surd$ & $\surd$ & $\surd$ & $\surd$ &\\
  W\"ustite & FeO                                             & $\surd$ & $\surd$ & - & - & & $\surd$ & $\surd$ & - & - & \\
  Magnetite & Fe$_3$O$_4$                                     & - & - & $\surd$ & - & & - & - & $\surd$ & - &\\
  Maghemite & Fe$_2$O$_3$                                     & - & - & - & $\surd$ & & - & - & - & $\surd$ & \\
  Troilite  & FeS                                             & $\surd$ & - & $\surd$ & $\surd$ & & $\surd$ & - & $\surd$ & $\surd$ &\\
  Pyrite    & FeS$_2$                                         & - & $\surd$ & - & - & & - & $\surd$ & - & - &\\
  \hline
  \hline
  \end{tabular}
  \end{center}
  \end{table*}

Another step forward was to include low-metallicity systems, such as damped Lyman-$\alpha$ absorbers \citep[DLAs, see, e.g.,][]{Wolfe05}, to study dust-depletion sequences \citep[][henceforth, Paper I]{DeCia16}. This is more complex than in the Galaxy because the abundances in DLAs can vary due to intrinsically low metallicity, nucleosynthesis effects, and dust depletion. Nevertheless, this degeneracy was bypassed by referring to the relative abundances (rather than abundances) of several metals, with different refractory and nucleosynthetic properties. The dust depletions were presented as a function of the observed relative abundance of Zn with respect to Fe, which is a dust tracer. More precesely, we use [Zn/Fe] $= \log N({\rm Zn})/N({\rm Fe}) - N({\rm Zn})_\odot /N({\rm Fe})_\odot$, where $N$ are the column densities and [Zn/Fe] $=0$ corresponds to solar abundance. From the dust depletion, \citet{DeCia16} derived elemental abundances in dust, which we use in this paper to infer the composition of dust, from low-metallicity systems to the Galaxy. We note that DLAs trace diffuse neutral gas with density usually of the order of $n \sim 1 - 10$~cm$^{-3}$.

Here we use a simplified model of the dust composition to reproduce the elemental abundances in dust from observations of the Galaxy and DLAs. We include dust species such as silicates (olivine and pyroxene series), iron-bearing oxides, and sulfides. Carbonaceous dust is not considered since data on carbon depletion is scarce.

\section{Data}
In this paper we use the elemental abundances in dust $\epsilon_X$ described in Paper I. These abundances are derived from a large (74) sample of DLAs towards quasars, all observed at high spectral resolution with the Very Large Telescope (VLT) Ultraviolet and Visual Echelle Spectrograph (UVES). In addition, the abundances of Galactic clouds from the HST observations of \citet{Jenkins09} were included. The $\epsilon_X$ are defined as
\begin{equation}
\epsilon_X = \log  \left\{  (1 - 10^{\delta_X}) \times 10^{\left[  [X/{\rm H}]_{\rm tot} + \log \left( \frac{N(X)}{N({\rm H})} \right)_\odot \right]}  \right\} \mbox{,}
\label{eq epsilon}
\end{equation}
where $\delta_X$ is the depletion of an element $X$ into dust grains, in a log scale (i.e., the atoms that are missing from the gas-phase with respect to the total abundance $\delta_X=[X/{\rm H}]-[X/{\rm H}]_{\rm tot}$). $[X/{\rm H}]_{\rm tot}$ is the total (dust-corrected) abundance of an element $X$ in each system, as described in Paper I. 

The $\epsilon_X$ describes the number of atoms of $X$, per hydrogen atom, that build up dust grains; these vary with the depletion-factor [Zn/Fe]. We report the elemental abundances in dust from Paper I in Table \ref{tabdata} for completeness. The typical value of [Zn/Fe] is low  for DLAs (never above [Zn/Fe]~$=1.0$), while it is in the range 0.7 to 1.6 for the Galaxy.

\section{Modeling the dust composition}

\subsection{Dust condensation}
\label{chemistry}
In this paper we try to distinguish the main components of the dust composition that could reproduce the observed elemental abundances, without assumptions on the process(es) that actually form the dust. Physically, the formation of dust in the ISM is through condensation upon existing seed grains (produced by stars). There are two different types of condensation that can occur in space: condensation of ices (volatiles) from molecules such as H$_2$O and CO$_2$, which form ice mantles on the seed grains, and growth of grains by chemical reactions on the surface of the grains when hit by certain molecular ``growth species''. The latter form of condensation corresponds to the formation of conventional dust, that is, the formation of solid-state matter in the form of minerals or carbonaceous material, which can survive on a long timescale. However, both types of condensation lead to depletion of certain elements, although not necessarily the same elements.

In the present paper we are mainly interested in dust depletion that stems from the chemical grain growth in the ISM. The densities needed to obtain efficient growth are typically found in cold molecular clouds. 

In the data set provided in Paper I, there is no information about carbon, simply because the carbon lines are typically saturated and in most cases could not be measured properly. Not knowing the level of carbon depletion, we are constrained to consider dust chemistry in the oxygen-rich regime. Therefore, our analysis is limited to silicates, oxides, and other carbon-free species. Such a dichotomy of carbon-rich and oxygen-rich chemistry is possible because very few dust species in these two regimes are formed from growth species that are important also in the other regime, that is, the molecules involved in the formation of silicates and various refractory-element oxides are not the same as those involved in the formation of carbonaceous dust. The only ``cross-over species'' worth mentioning is SiC, which in principle could have some bearing on the depletion of both silicon and carbon. The interstellar abundance of SiC is relatively low however \citep{Whittet90}.

\subsection{Dust chemistry matrix}
\label{chematrix}

In the following we attempt to identify the statistically preferred composition of dust, given the abundances in dust and their uncertainties derived in Paper I. The silicate-dust complex likely contains a mixture of different dust species. It seems, however, that the majority of silicate dust is dominated by two types: olivines and pyroxenes \citep{Draine03}. Asymptotic giant branch stars probably produce mainly almost iron-free olivine (forsterite) and/or pyroxene (enstatite) grains \citep{Hofner08, Norris12, Bladh12}. It is likely that silicates produced in supernovae/high-mass stars are also iron-free, because iron-free silicates are less susceptible to radiative heating and sublimation \citep[see, e.g., results by][]{Todini01}. Also, observed SN-produced silicates seem to be iron-poor, because SN-produced dust has been observed mostly for core-collapse SNe \citep[see, e.g.,][]{Matsuura11}, but almost never for Type Ia SNe \citep[see][for a rare exception]{Nagao18}, which are the major iron producers.

Nonstoichiometric compounds such as olivine, pyroxene, and w\"ustite can be described in terms of the ``rule of mixtures'', where a parameter $x$ in the interval $0 \le x \le1$ regulates the composition.  For olivine and pyroxene we use the composition parameters $x_{\rm ol}$ and $x_{\rm py}$ as model parameters, which measure the iron content of silicates, albeit with some uncertainty. Because $x_{\rm ol}$ and $x_{\rm py}$ are treated as free parameters, we compute model grids covering all possible $x_{\rm ol}$ and $x_{\rm py}$ values. In principle we could also include the composition parameter for Fe$_{1-x_{\rm w}}$O, because the FeO stoichiometric phase is never stable. Because this compound is still close to stoichiometric, and the composition varies very little, we do not consider the composition parameter of w\"ustite a model parameter and simply assume the FeO stoichiometric phase. We assume the same for sulfides -- either we use the stoichiometric phase FeS (troilite) or FeS$_2$ (pyrite). In addition, we also consider Fe$_3$O$_4$ (magnetite) and Fe$_2$O$_3$ (hematite/maghemite) as alternatives to FeO.

Our composition model has five more parameters which represent the number abundances of the considered elements (O, S, Mg, Si, Fe) in the assumed dust species. For a given set of stoichiometry parameters, these elemental abundances in dust are relatively well-constrained by the  observed depletion pattern. For a given pair of stoichiometric parameter values, we can calculate the expected solid-phase elemental abundance vector $\bf{D}$ (the abundances in dust grains of the considered elements) from
\begin{equation}
\label{matrix}
\mathbf{D} = \mathbf{X}\,\mathbf{W}
,\end{equation}
in which the {\it dust chemistry matrix} $\mathbf{X}$ is given by
\begin{equation}
\mathbf{X} = \left[
\begin{array}{lllll}
4 & 3 & 0 & b & 0 \\
2\,x_{\rm ol} & x_{\rm py} & 0 & 0 & 0 \\
1 & 1 & 0 & 0 & 0 \\
0 & 0 & 0 & 0 & c \\
2\,(1-x_{\rm ol}) & 1-x_{\rm py} & 1 & a & 1\\
\end{array}
\right],
\end{equation}
where the parameters $a$, $b,$ and $c$ define the four composition types that we have chosen to consider. Namely, $a = 1,2,3$, $b = 1,3,4$ depending on whether the iron oxide is FeO, Fe$_2$O$_3,$ or Fe$_3$O$_4$ and $c = 1,2$, depending on whether the iron sulfide is troilite (FeS) or pyrite (FeS$_2$). The resultant vector $\mathbf{W}$ contains the number abundances of each considered dust species.

All four compositions A1 -- D1 include olivine, pyroxene, and metallic iron to account for silicon, magnesium, and iron depletion. Type A  also includes FeS to account for sulfur depletion and FeO to account for the generally high abundances of oxygen and iron in dust. Type B is identical to A, except that FeS is replaced with FeS$_2$. Types C and D are similar to A, but the iron oxide is Fe$_3$O$_4$ or Fe$_2$O$_3$, respectively, instead of FeO. For an overview of the composition types, see Table \ref{parameters}. 

In addition we also consider four modified compositions, A2 -- D2, where the silicate component is assumed to consist of both iron-free as well as iron-rich silicates and the amount of sulfur in dust is reduced. The reasons for these modifications and properties of these composition types are explained in Sect. \ref{modified}.

The problem of constructing a composition model based on the species we have discussed above is now equivalent to finding the vector $\mathbf{W}$ and composition parameters $x_{\rm ol}$ and $x_{\rm py}$  (specifying the chemistry matrix $\mathbf{X}$) which gives the smallest discrepancy between the observationally inferred abundances in dust grains and the elemental abundance vector $\mathbf{D}$. 

The above approach is limited to composition models where the number of dust species is the same as the number of depleted elements considered in the model. Moreover, two dust species that consist of exactly the same elements (such as pyroxene and olivine, or FeO and Fe$_3$O$_4$) add some degree of redundancy and degeneracy to the model, since two similar dust species will constrain the depletion pattern less than two chemically very different species (e.g., enstatite and FeS). While these are indeed limitations of our simplified model, we note that as long as the dust species dominating the dust mass budget are included among the dust species that we consider in our model, at least in some of the composition types, this basic approach should provide a first approximation of the composition properties, as well as solid indications of the overall trends in the dust mass budget.

\subsection{Monte Carlo simulation}

The resultant vector $\mathbf{W}$ contains number abundances of each considered dust species, for a given elemental abundance vector $\mathbf{D}$. However, numerically these values are necessarily positive. This is why we have to use a statistical approach and take the variance of each elemental abundance in dust into account. In addition, a statistical (Monte Carlo) treatment can provide a better handle on the formal uncertainties of the results of our model, despite its simplified nature.

We model the dust composition stochastically, assuming that the $1\,\sigma$ errors given in Table \ref{tabdata} correspond to the standard deviation of a Gaussian error distribution (truncated at $\pm 3\,\sigma$), that is, we generate a set of abundances, for each element $i$, by randomly drawing a deviation $\Delta\epsilon_i$ from a Gaussian distribution with a $\sigma$ corresponding to the values given in Table \ref{tabdata} and adding this deviation to the observationally inferred abundances in dust $\epsilon_i$. The random $\Delta\epsilon_i$ values are generated using the Box-Muller transform \citep{Box58} combined with a standard (uniform) random number generator. The abundances $\epsilon_i + \Delta \epsilon_i$ are then used to solve the matrix equation given in Section \ref{chematrix} for a given combination of the parameters $x_{\rm ol}$, $x_{\rm py}$, $a$, $b$ and $c$, in order to obtain abundances of the considered dust species.

The randomisation procedure described above is repeated a large number of times ($10^7$) and from the resultant set of ``mock compositions'', we extract those which do not generate negative dust abundances; that is, after solving the matrix equation (\ref{matrix}) for each randomised $\bf{D}$, we disregard all solutions for which there are negative entries in $\bf{W}$. It must be emphasized that we are thus only keeping a small fraction of all solutions generated. This is done so that we can limit the possible solutions within the parameter space that is physically meaningful. The final set of compositions is reduced to a set of a few times $10^4$ solutions, which is less than 1\%\ of the total number of solutions, but is still large enough to obtain statistically significant expectation values for the number abundances of each dust species. The matrix inversion is done using an LU decomposition algorithm \citep{Press92}.

Since we have 4+4 different composition types (see Table \ref{parameters}) and two essentially unknown parameters $x_{\rm ol}$ and $x_{\rm py}$ regulating the amount of iron in silicates, we have to compute a large grid of Monte Carlo simulations in order to cover the full parameter space. For each composition type A1 -- D1 and A2 -- D2 specified in Table \ref{parameters}, we compute grids of models, covering the $x_{\rm ol}$ -- $x_{\rm py}$ plane in steps of $0.05$; that is, for each composition type we perform 441 Monte Carlo runs (each based on $10^7$ patterns of elemental abundances in dust with randomised Gaussian deviations as described above) corresponding to different iron contents of the two silicate species (olivine and pyroxene). In total we try $1.764\cdot 10^{10}$ randomised depletion patterns, which in combination with numerical matrix inversion requires computing power that is moderate but not insignificant. For a standard quad core Intel processor ($\sim 2\,$KHz) the processing time is roughly 200 hours.

\subsection{Modified silicate composition (A2 -- D2)}
\label{modified}
For each of the composition types (A1 -- D1) there are a few implicit assumptions. First, it is assumed that there exists only two representative silicate species, that is, that we can model the silicate component as olivine and pyroxene with representative values for $x_{\rm ol}$ and $x_{\rm py}$.  This is not necessarily a good model, but serves as a reasonable ``first-order approximation'' of reality since, effectively, there is an overall iron-magnesium-abundance ratio that can be associated with representative values of $x_{\rm ol}$ and $x_{\rm py}$. Second, sulfur is a problematic element \citep[see the discussion by][]{Jenkins09} and the amount of it that goes into sulfides is not necessarily similar to the abundance of sulfur that is not in the gas phase. 

Composition types (A2 -- D2) represent an attempt to relax these assumptions. Instead of single values of $x_{\rm ol}$ and $x_{\rm py}$, we consider the possibility of having dust grains that coexist with two different values of $x_{\rm ol}$ and $x_{\rm py}$, respectively. Assuming that 50\% (by number) of the silicate grains are iron-free (forsterite and enstatite, which are end members of the olivine and pyroxene sequences, respectively) and that the remaining 50\% are iron-rich olivine and pyroxene with representative values for $x_{\rm ol}$ and $x_{\rm py}$, we can obtain a different magnesium-to-iron ratio. Similarly, we also make the {\it ad hoc} assumption that only 25\% of the ``missing'' sulfur is found in sulfides.

\subsection{Mass fractions}
The output of the final Monte Carlo simulations is a set of possible dust compositions in terms of normalised number abundances. These number abundances can easily be transformed into mass fractions $\mu_i$ of the total (noncarbonaceous) dust component by using the effective mass numbers $A_i$ for each dust species $i$ (e.g., $A_{\rm FeO} = A_{\rm Fe} + A_{\rm O}$ = 71.84). We then compute the expectation values,
\begin{equation}
E(\mu_i) = {1\over N}\sum_{j=0}^{N-1} \mu_i,
\end{equation}
where $N$ is the number of plausible compositions generated by the Monte Carlo code. Similarly, we also compute the variance as
\begin{equation}
{\rm Var}(\mu_i) = {1\over N-1}\sum_{j=0}^{N-1} \left[\mu_i - E(\mu_i)\right]^2.
\end{equation}
Since we compute a grid, trying to cover the $x_{\rm ol}$ -- $x_{\rm py}$ plane, we obtain also a distribution of expectation values for the mass fractions. This distribution appears to be close to Gaussian with a relatively small variance. Thus, we calculate the mean mass fractions (statistically speaking, the ``expected expectation values'') for each composition type, assuming that $x_{\rm ol}$ and $x_{\rm py}$ are completely arbitrary (which is not entirely true), and use these values as estimates of the most likely combination of mass fractions of different dust species.
The range of expected abundances  of each depleted element in dust is estimated from the $1\,\sigma$ deviations from the ``expected expectation values'' for the mass fractions.

    \begin{figure*}
    \resizebox{\hsize}{!}{
  \includegraphics{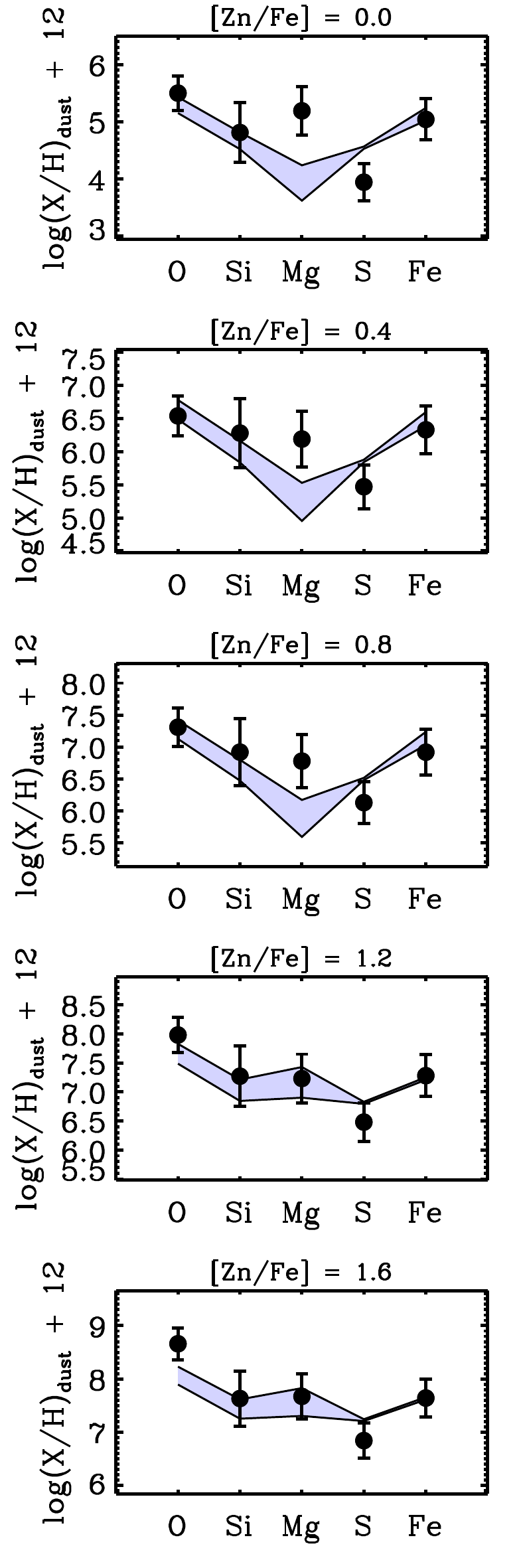}
  \includegraphics{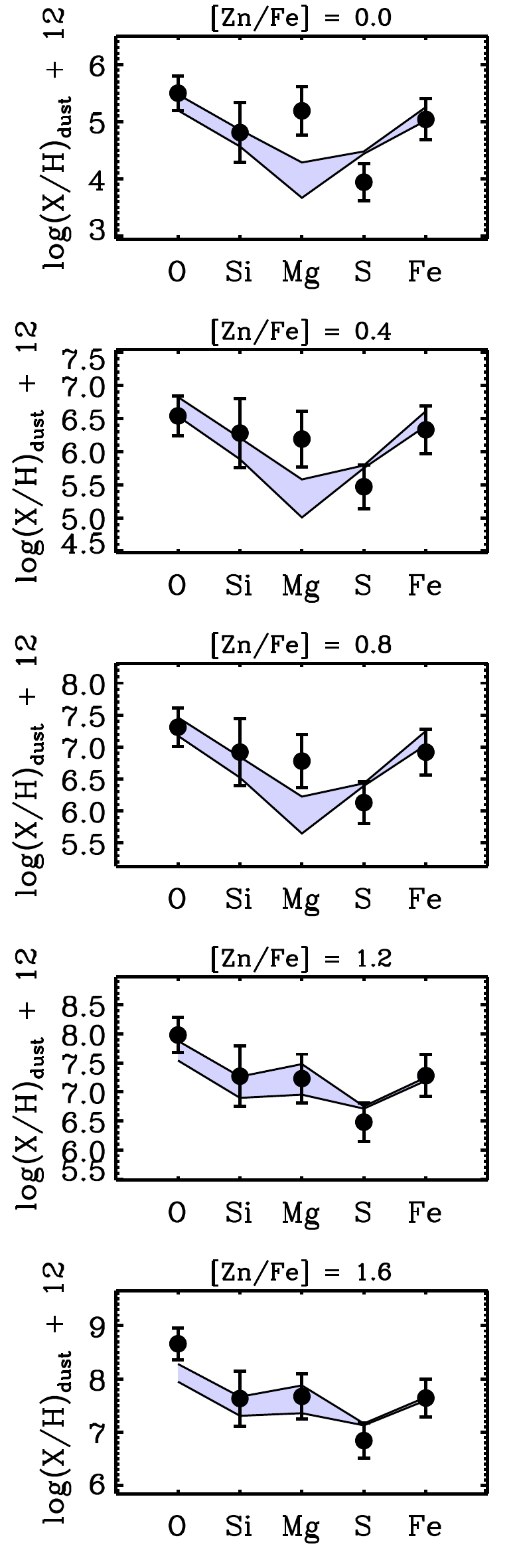}
  \includegraphics{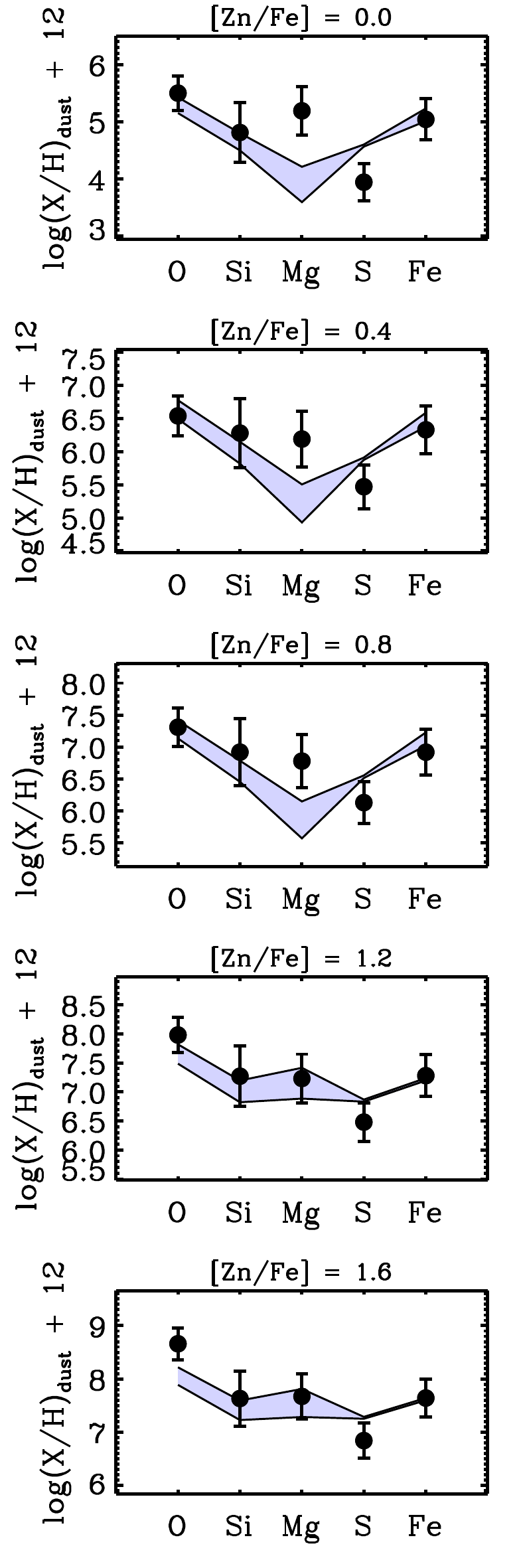}
    \includegraphics{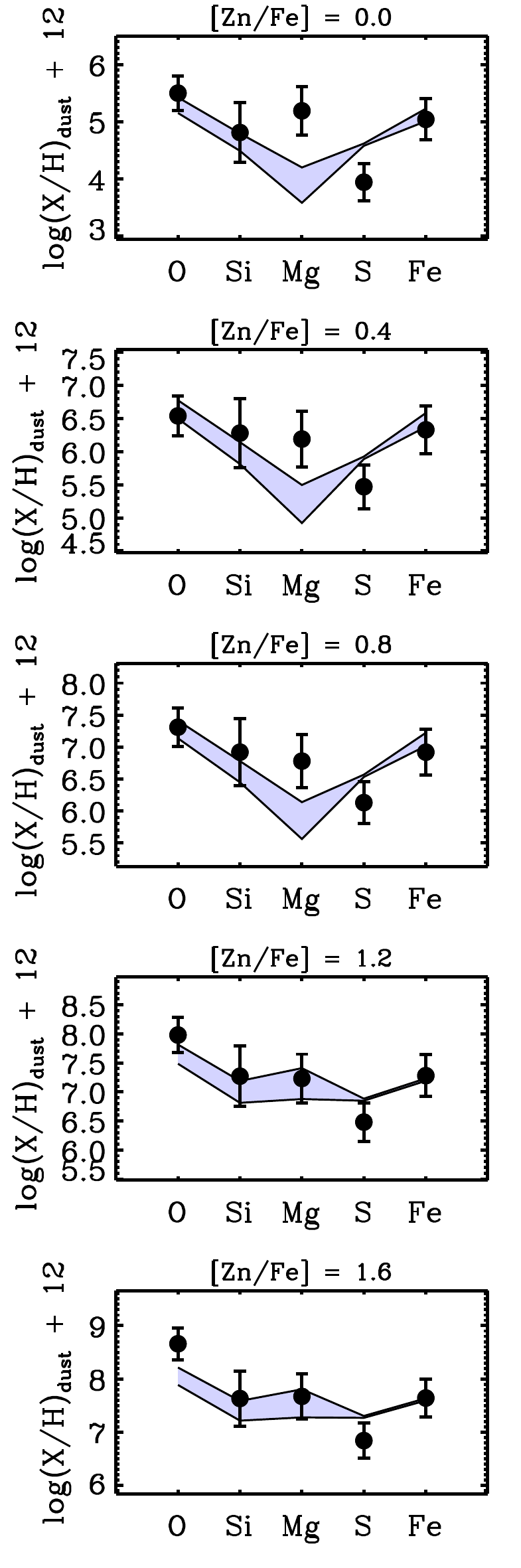}}
  \caption{  \label{abuindust}
   The abundance of the five considered elements in dust for different [Zn/Fe]. The different columns represent composition type A1, B1, C1, and D1 (from left to right). The shaded regions show the $\pm 1\,\sigma$ range of the Monte Carlo models and the filled black circles with error bars represent the data from Paper I.  }
  \end{figure*}
  
      \begin{figure*}
    \resizebox{\hsize}{!}{
  \includegraphics{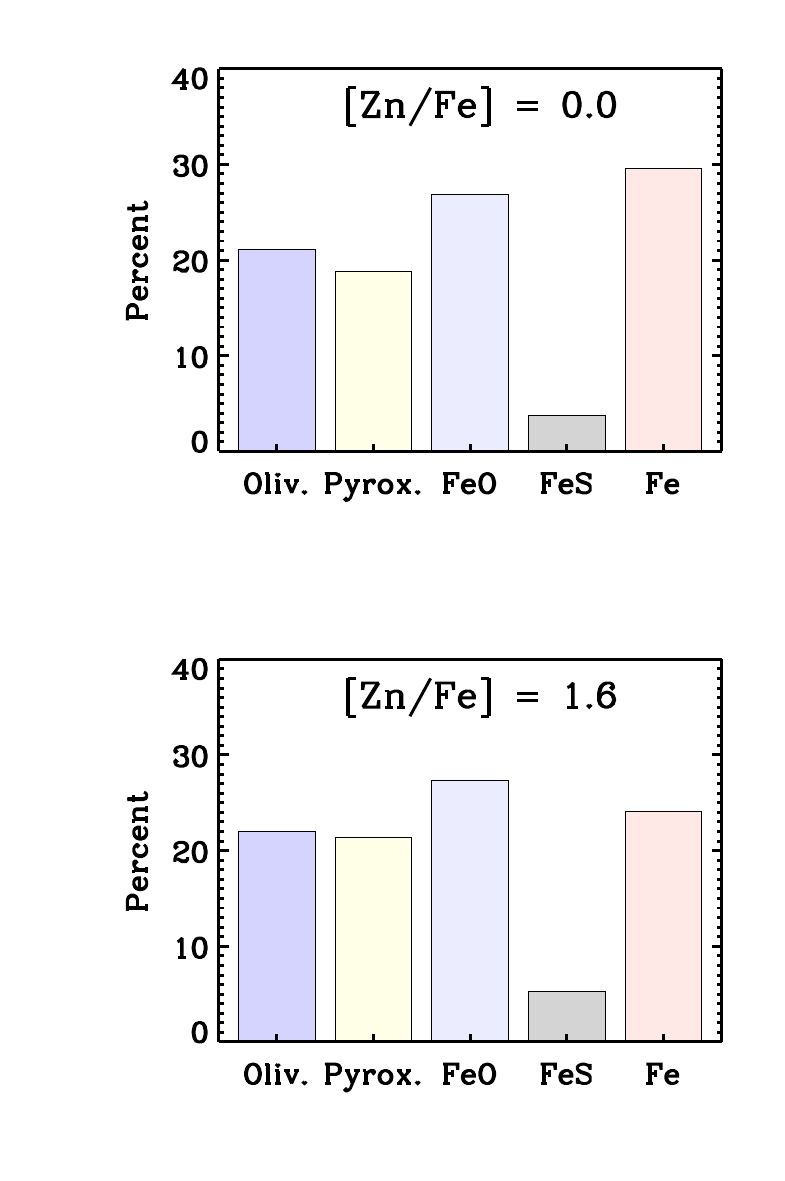}
  \includegraphics{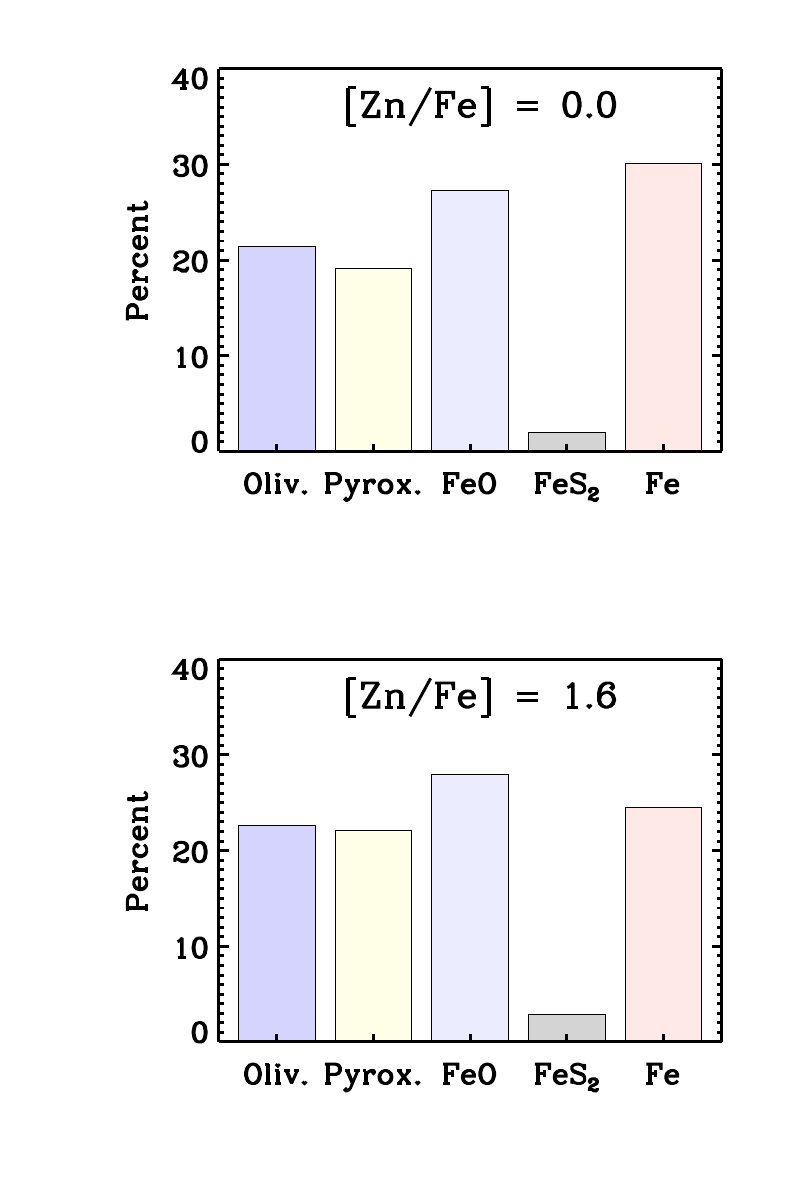}
  \includegraphics{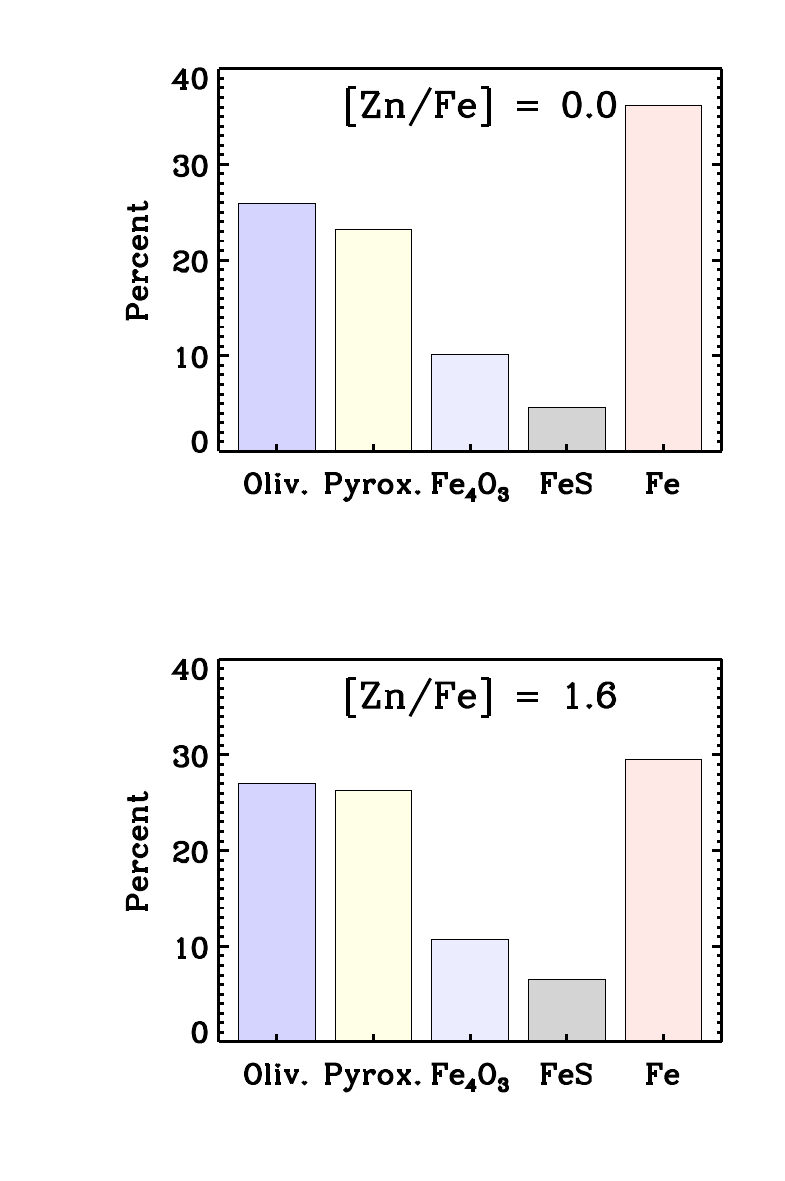}
    \includegraphics{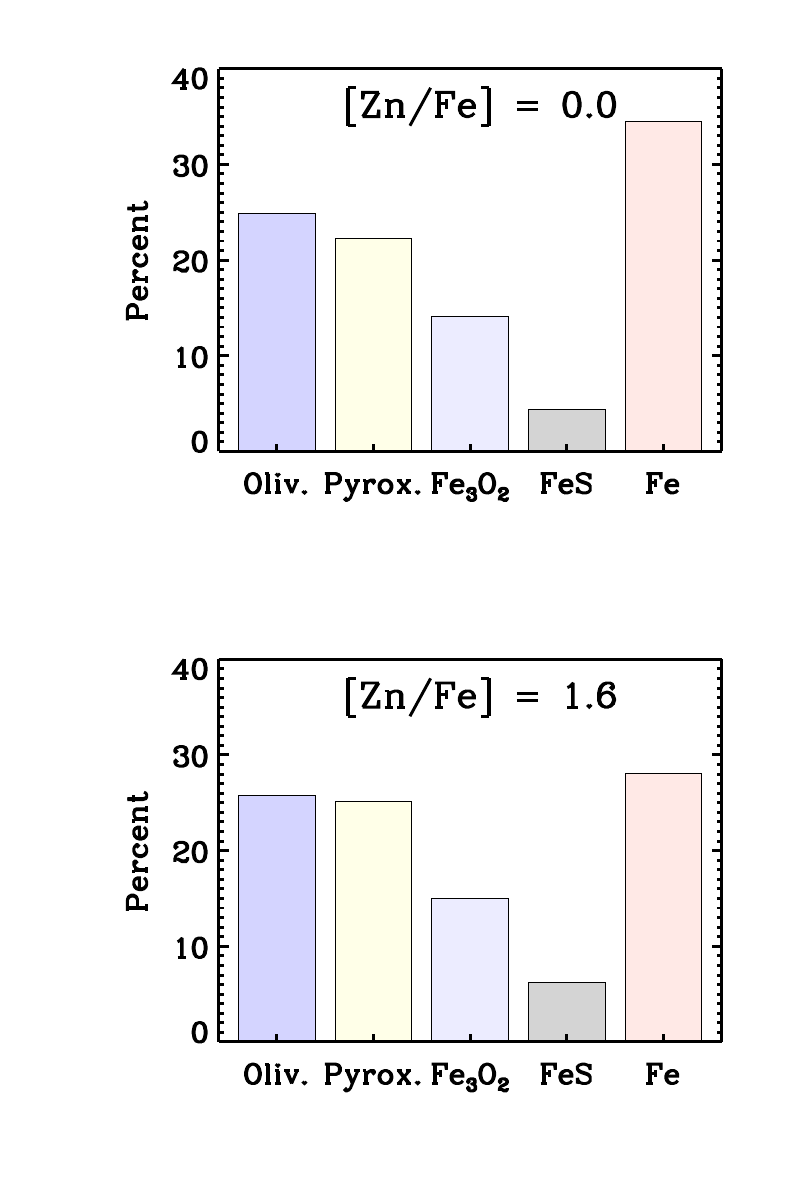}}
  \caption{  \label{massfractions}
 Mass fractions of each dust species in the considered composition types A1, B1, C1, and D1 (from left to right) for the low-depletion regime ([Zn/Fe] = 0) and for the Galactic level of depletion ([Zn/Fe] = 1.6).}
  \end{figure*}

      \begin{figure*}
    \resizebox{\hsize}{!}{
  \includegraphics{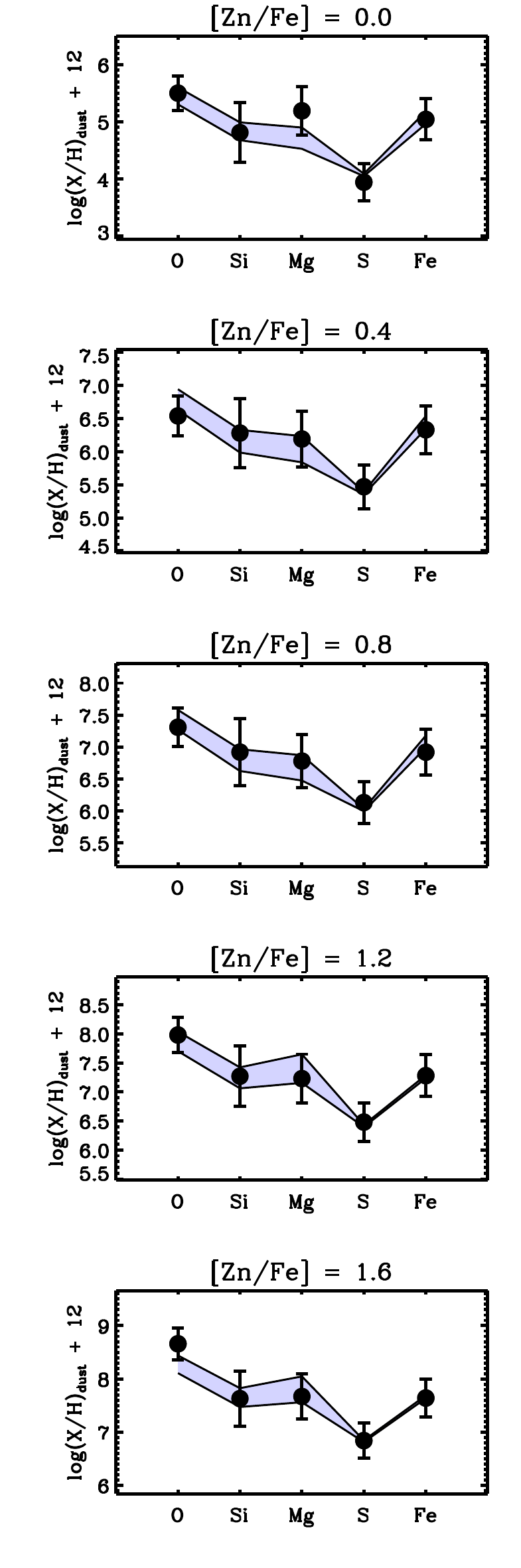}
  \includegraphics{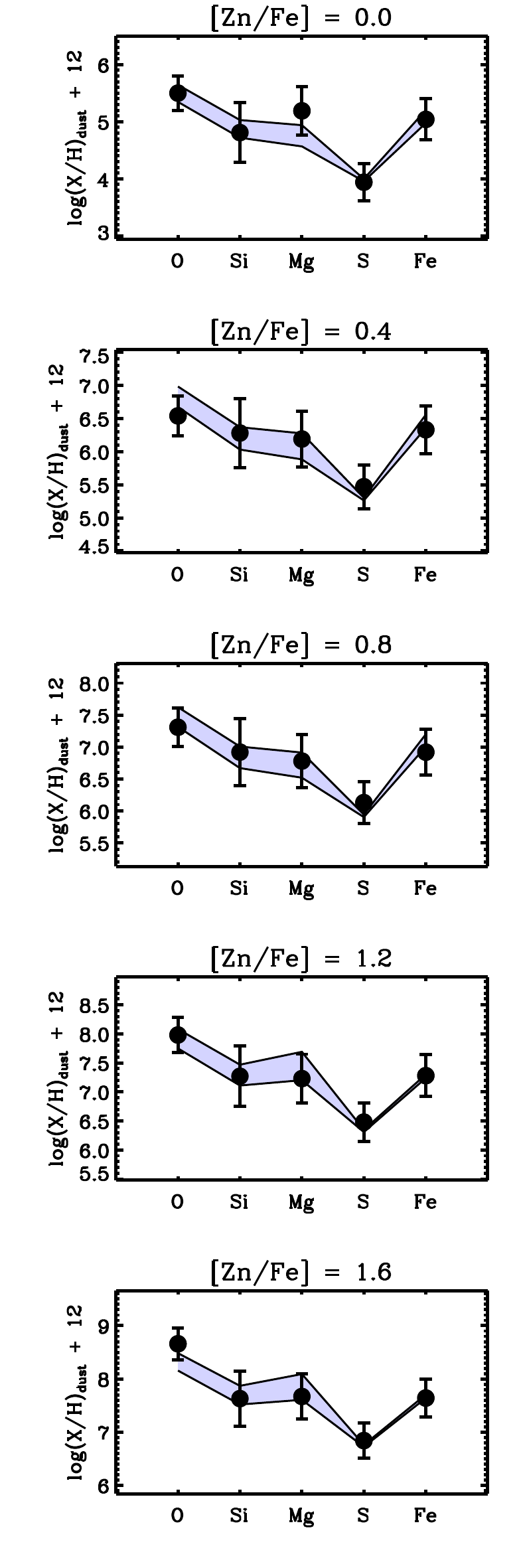}
  \includegraphics{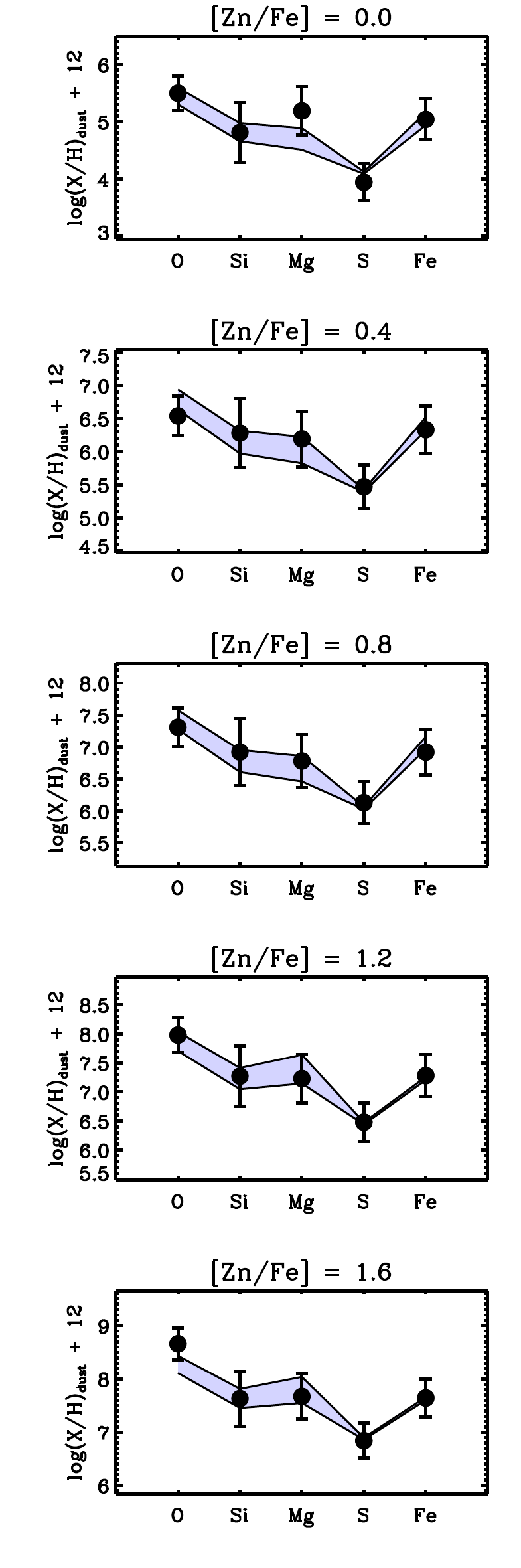}
    \includegraphics{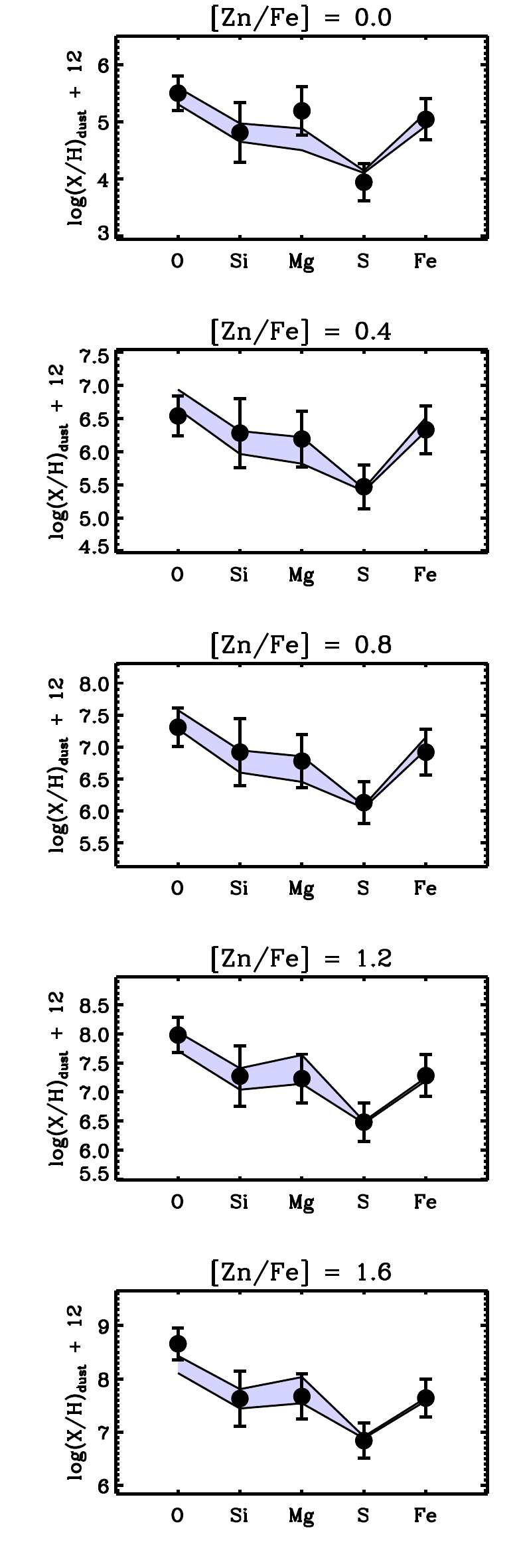}}
  \caption{As in Fig. \ref{abuindust}, but for the modified compositions A2 -- D2 described in Section \ref{modified}.  \label{abuindust_mod}
   }
  \end{figure*}
  
        \begin{figure*}
    \resizebox{\hsize}{!}{
  \includegraphics{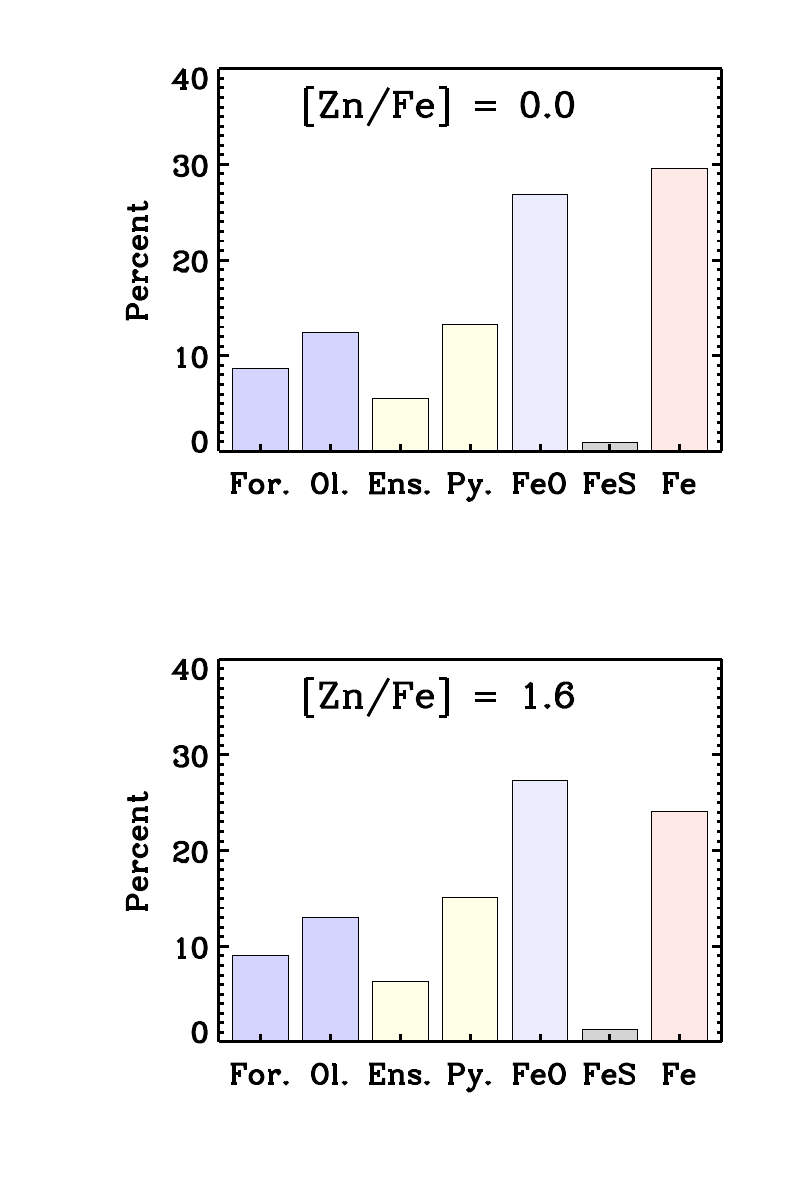}
  \includegraphics{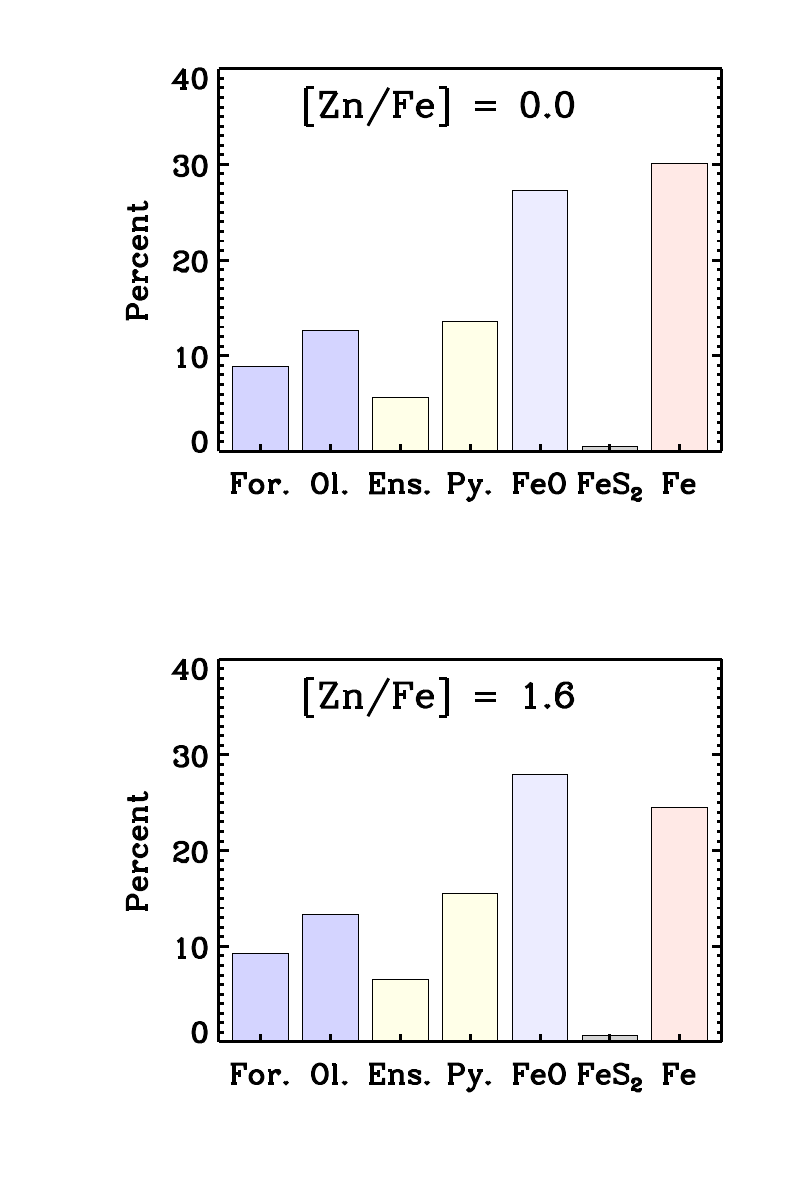}
  \includegraphics{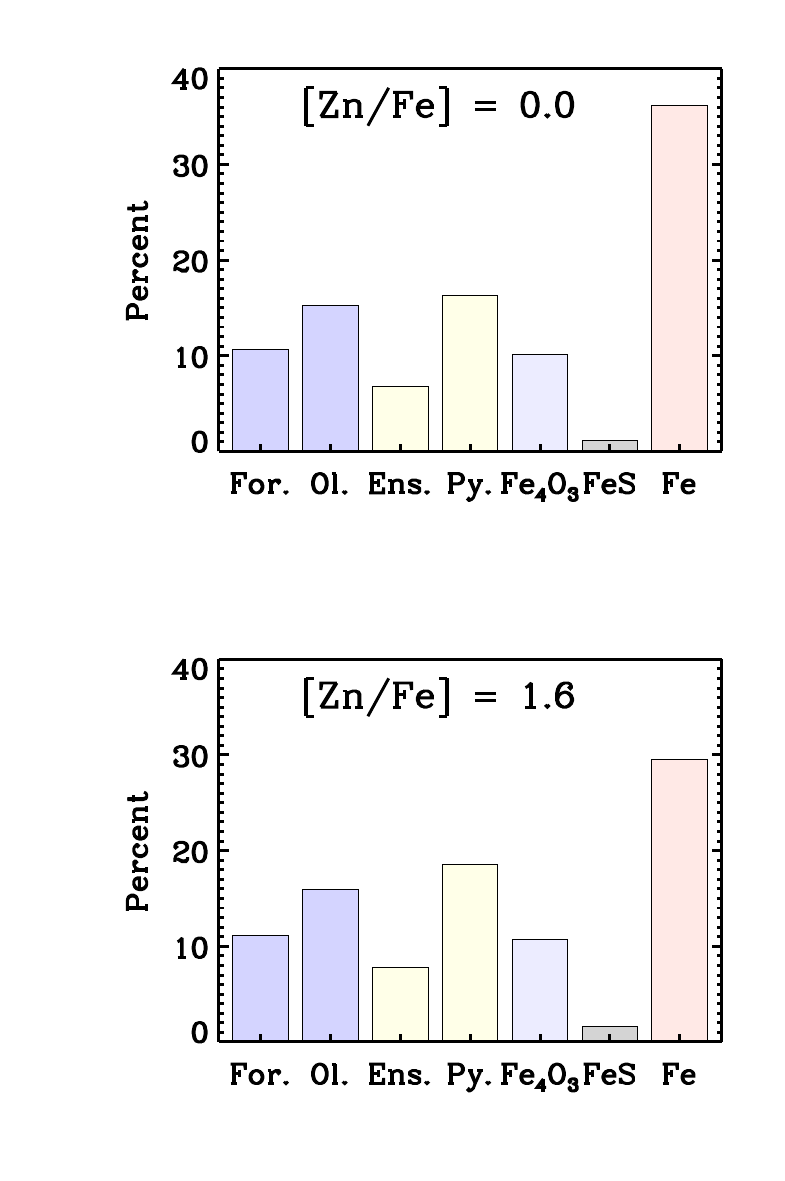}
    \includegraphics{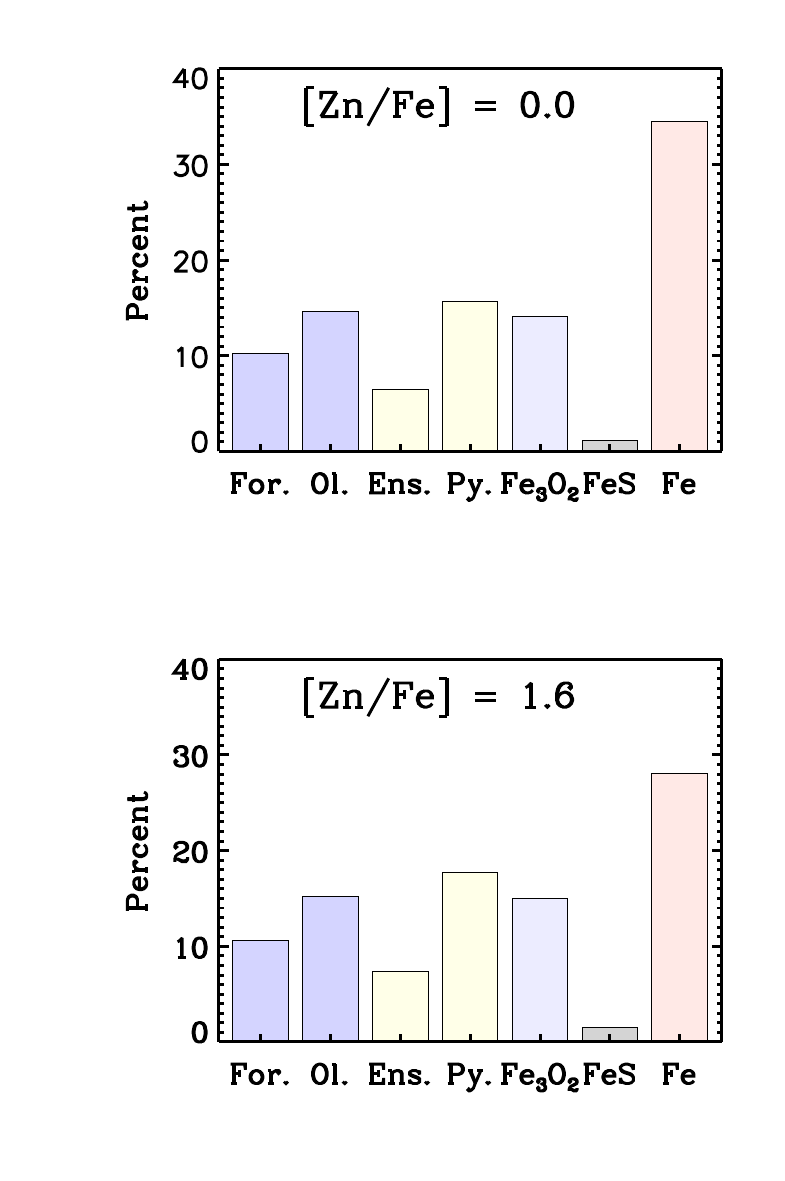}}
  \caption{ As in Fig. \ref{massfractions}, but for the modified compositions A2 -- D2 described in Section \ref{modified}.  \label{massfractions_mod}
   }
  \end{figure*}

\section{Results}
\label{results}
In Fig. \ref{abuindust} we show the overall agreement between the models and the observed abundances in dust. Composition types A1 to D1 are shown in columns from left to right. Each row of panels corresponds to a given [Zn/Fe], that is, to increasing levels of dust content. The shaded areas show the $\pm 1\,\sigma$ range and the filled black circles with error bars represent the data from Paper I (also Table \ref{tabdata}). Clearly, the agreement between models and data is better at higher degrees of depletion. However, the modeled oxygen abundance at [Zn/Fe]~$=1.6$ is not within the error bar of the observed abundance. All four A1 -- D1 composition types seems to be in disagreement with the observed magnesium and sulfur at [Zn/Fe]~$=0.0$. The modeled abundance of sulfur is at best only marginally consistent with the observed values, regardless of the level of depletion. Modeled magnesium abundances agree well with the observed values for [Zn/Fe]~$=1.2$ and [Zn/Fe]~$=1.6$, however.

In general, there are only small differences between the predicted depletion patterns for the four A1 -- D1 composition types. There is not one among them that is clearly preferred, although type B1 appears to generate a marginally better fit to the data. The corresponding mass fractions of each dust species are shown in Fig. \ref{massfractions}, for the lowest and highest values of [Zn/Fe] only. 

There is very little evolution in the mass fractions from [Zn/Fe]~$=0.0$ to [Zn/Fe]~$=1.6$ (Fig. \ref{massfractions}). Between the different composition types A1 -- D1 there are some differences in the mass fractions of iron oxides. Overall, however, the expected distribution of the dust mass into silicates, iron oxide, iron sulfide and metallic iron is quite similar for all levels of depletion and all composition types. 

Regardless of the composition type assumed, all models seem to indicate a slight increase of the pyroxene fraction with increasing [Zn/Fe]. The abundance of magnesium in dust increases with [Zn/Fe], which is probably the explanation for this trend. We also see that, in all models, the noncarbonaceous dust that we consider is rich in ``iron dust'', such as iron oxides and metallic iron grains, which actually seems to dominate the dust-mass budget (but not necessarily by number). 

Composition types A2 -- D2 represent \textit{ad hoc} modifications that we introduce to address the shortcomings of the A1 -- D1 models, that is, the disagreement between observations and model predictions for the oxygen abundances at [Zn/Fe]~$=1.6$, and magnesium and sulfur abundances at [Zn/Fe]~$=0.0$. On one hand, depletion of oxygen into molecules may play a role, as we discuss in Sect. \ref{illconstrained}. On the other hand, the effective iron content of silicates may not be adequately modeled with the individual values for the parameters $x_{\rm ol}$ and $x_{\rm py}$ assumed in composition types A1 -- D1. In fact, the interstellar dust component may be a mixture of silicates with various abundances of iron in them. Most notably, one can expect the coexistence of both iron-rich and essentially iron-free silicates (forsterite and enstatite). The latter are likely of stellar origin present in the ISM, either as newly formed stellar dust or as the cores of silicates that have accreted a mantle of more iron-rich material. Sulfur may also be less depleted than the models suggest. Overall, we find a good agreement between the modified A2 -- D2 models and the observed data. 

The mass fractions of olivine and pyroxene in compositions A1 and A2 as a function of the silicate composition parameters $x_{\rm ol}$ and $x_{\rm py}$  (derived from the grid of models) show relatively little variation over most of this parameter plane. Figures \ref{modelA_ol} and \ref{modelA_py} in Appendix A show this graphically. The only exception is the region corresponding to relatively small  $x_{\rm ol}$ and $x_{\rm py}$ values, more precisely $x_{\rm ol}\lesssim 0.2$, $x_{\rm py}\lesssim 0.1$, in which there is a steep rise of the mass fraction of olivine and pyroxene (see Figs. \ref{modelA_ol} and \ref{modelA_py}). It is worth noting that the two types of silicates are a bit like ``mirror images'' of each other in terms where the highest mass fractions occur in the $x_{\rm ol}$ -- $x_{\rm py}$ plane. For olivine, this ``peak'' is found along the $x_{\rm ol}$-axis and, similarly, for pyroxene it is found along the $x_{\rm py}$-axis.

\section{Discussion}
\label{discussion}

\subsection{Which is the best-fit composition?}

Despite the limitations of our simplified approach and the degeneracy of some parameters, from the results of the different models we can robustly conclude that: (1) the dust composition does not change dramatically as a function of the overall dust content, from DLAs to the Galaxy; (2) the noncarbonaceous dust mass is dominated by ``iron dust'', that is, iron oxides and metallic iron, which make up at least half of the dust mass; and (3) there must be both iron-poor and iron rich silicates present at all [Zn/Fe], although the relative proportions are difficult to decipher with only depletion data. The previous consensus on a picture of interstellar dust as mainly a mix of ``astronomical'' iron-rich silicates and carbonaceous dust/PAHs \citep[see, e.g.,][]{Draine84,Draine03,Draine07,Pei92,Weingartner01} must however be put into question once again.

The modified dust composition types (A2 -- D2) assume that 50\% of the olivine and pyroxene grains belong to the iron-free end-members forsterite and enstatite, respectively, while the remaining 50\% are iron-rich with representative values for $x_{\rm ol}$ and $x_{\rm py}$. These modified compositions yield good fits to the data (Fig. \ref{abuindust_mod}) where the unmodified compositions (A1 -- D1, Fig. \ref{abuindust}) do not. But the {\it ad hoc} modifications we made neither necessarily reflect what the best-fit composition should be, nor do they reflect what the true composition is. However, it seems clear that the silicate component is an admixture of iron-rich and iron-poor silicates. A mounting body of evidence \citep[see, e.g.,][]{Min07,Jones07,Altobelli16,Sofia06} suggests magnesium-rich (iron-poor) silicates could dominate the interstellar zoo of silicates and therefore increase the need for other iron-bearing species. However, we present only the 50/50 modification here, as it provides the best fit to the depletion data. It should be noted however that a test model with iron-free silicates was also found to fit the data relatively well. It is also clear that depletion into molecular gas, or possibly in the form of ice mantles, must play a role, at least for oxygen (see Section \ref{illconstrained} for further discussion). 

If the ``missing iron'' cannot be in silicates, as in the SGM case,  but rather in iron oxides and metallic iron, we must consider the implications of including very significant amounts of such iron-bearing species. Spherical grains of metallic iron add a featureless power-law component to the extinction curve, ranging from the optical out to the far infrared. Iron oxides, on the other hand, show features that could affect the extinction curve such that it may be possible to talk about a detection of extinction by iron oxides. W\"ustite (FeO) shows a wide absorption feature starting at approximately $20\,\mu$m and extending out to $\sim 30\,\mu$m that is somewhat dependent on the assumed grain properties \citep{Andersen06}. Maghemite ($\gamma$-Fe$_2$O$_3$) shows a strong feature around $11\,\mu$m, and a double peaked feature due to magnetite (Fe$_3$O$_4$) may also show up at a similar wavelengths \citep[see Fig. 6 in][]{Draine12}. Unfortunately, this is very close to the $10\,\mu$m feature expected from silicates and since the $10\,\mu$m feature can be relatively wide, it can sometimes be difficult to distinguish between maghemite, magnetite, and iron-bearing silicates. A relatively high abundance of iron oxides is therefore not at all unlikely, but whether something like a quarter of the total dust mass is realistic or not requires detailed analysis of extinction properties. We therefore postpone all further discussion of the best-fit composition to our forthcoming theoretical study of extinction curves \citep[][henceforth Paper III]{PaperIII}. 

\subsection{The onset of interstellar dust condensation}  
The depletion patterns derived observationally (see Paper I) suggest that silicon, magnesium, sulfur, and iron are depleted in similar proportions for each step in [Zn/Fe], although the absolute level of depletion changes. Silicon, magnesium, and iron all have relatively similar characteristic condensation temperatures \citep[see, e.g., Fig, 15 in][]{Jenkins09}, which suggest that the onset of  depletion of these elements (relative to other elements) should occur at roughly the same time.

At zero metallicity, the two dominating dust species formed in
type II SNe are, according to some models, metallic iron and amorphous carbon \citep[see, e.g.,][]{Todini01}. The bulk of silicates should then have formed somewhat later, at higher metallicity, predominantly by interstellar dust condensation. If this hypothesis is correct, that is, if the dominant channel of dust formation at later times is condensation in the ISM, the depletion of silicon and magnesium is also a marker of the onset of efficient interstellar dust condensation. The fact that our data from Paper I do not suggest any major shift in the proportions of silicon, magnesium, and iron may thus be indicative of interstellar dust condensation being the dominant dust-production channel at all levels of depletions and metallicities probed by the observations. 

\subsection{Silicon and magnesium: the well-constrained elements}
In Paper I we presented a set of data for silicon and magnesium, which covers a wide range of [Zn/Fe] values and has relatively small intrinsic scatter. The depletion of silicon and magnesium is adequately explained by the formation of a mixture of both iron-rich and iron-poor silicates (see Fig. \ref{abuindust_mod}), in agreement with previous studies of local interstellar environments \citep[][]{Jones87,Savage96b,Kimura03}. Other silicon- and magnesium-bearing species, such as silicon carbide (SiC), gehlenite (Al$_2$Ca$_2$SiO$_7$), spinel (MgAl$_2$O$_4$), diopside (MgCa(SiO$_3$)$_2$), and anorthite (CaAl$_2$Si$_2$O$_8$), to mention a few, are simply too rare \citep{Jones07} to make a significant contribution. It would be theoretically possible that the lack of magnesium for low [Zn/Fe] is explained by adding magnesioferrite (MgFe$_2$O$_4$), but there is no convincing evidence for the existence of this species in the ISM.

Provided that the {\it ad hoc} modification to the composition types we have considered is reasonable, we have a model which is in agreement with the depletion patterns at all considered levels of depletion ([Zn/Fe]). We find this to be conclusive evidence that silicon and magnesium are almost only depleted due silicate condensation, which is also consistent with previous work indicating that the Si/Mg depletion ratio is such that it could be explained with typical silicates \citep{Jones00}. However, this does not mean that a major fraction of the silicates cannot be iron-poor and magnesium-rich. A model with only iron-free silicates does, in fact, fit the data relatively well. Here we have nevertheless focused on demonstrating that SGM-type compositions do not adequately reproduce the depletion patterns from Paper I, while inclusion of 50\% (by number) iron-free silicates (compositions A2 - D2) seems to solve the problem and provides a better fit than 100\% iron-rich silicates (compositions A1 -D1) or 100\% iron-free silicates.

\subsection{Oxygen, sulfur, and iron: the not-so-well-constrained elements}
\label{illconstrained}
Our model has a tendency to underpredict the amount of oxygen in dust at high [Zn/Fe] (see Figs. \ref{abuindust} and \ref{abuindust_mod}), although the discrepancies only exceed the estimated observational uncertainty for composition types A1 -- D1 at [Zn/Fe]~$=1.6$. This indicates that there could be at least one component missing that has the ability to regulate the oxygen abundance in dust grains. Clearly, our model is incomplete in the sense that we have not considered the effect that molecular gas may have on the depletion pattern. For instance, oxygen could be depleted in oxygen-bearing molecules such as O$_2$, CO, CO$_2$, SiO, MgO and H$_2$O, which are expected to be more abundant at higher metallicities, in systems with higher levels of depletion. In particular, a non-negligible part of the oxygen can be locked up in carbon monoxide (CO), which is a molecule that is both abundant in molecular clouds and very hard to dissociate (it has the highest known dissociation energy for any molecule).  At high [Zn/Fe], CO could perhaps explain the overabundance of oxygen in dust \citep{Whittet10} compared to the models. However, the abundance of gas-phase molecules may not be enough to explain the ``missing-oxygen problem''. Therefore, it has been suggested that ices made of oxygen-bearing molecules could be the solution \citep{Poteet15}. Ice can readily condense on the surfaces of dust  grains to form ice mantles, which can easily be up to a 100 nm thick. There could therefore be plenty of ``hidden'' oxygen in the ISM simply locked up in these ice mantles.

Sulfur is not well-constrained for the Galaxy \citep[see][]{Jenkins09}, possibly because of ionisation effects in HII regions. The extent  to which sulfur is actually included in dust grains remains difficult to assess. The DLA data alone should not be affected by these ionisation effects, and do indeed suggest a depletion trend also for sulfur. Therefore, the depletion of sulfur is probably due to iron sulphates such as troilite (FeS) or pyrite (FeS$_2$), but since the abundance of metallic iron is an essentially free parameter, any iron sulphate could be used to explain the inferred sulfur abundance in dust.  Moreover, assuming that most of the sulfur is not in dust appears to give a better fit to the data (see Fig. \ref{abuindust_mod}) and is consistent with Galactic constraints on sulfur depletion \citep{Ueda05}. We caution the reader that the FeS or FeS$_2$ abundances implied by our models cannot be seen as well-constrained results.

As we mention immediately above (and also in previous sections), metallic iron is treated as an almost free parameter. Therefore, all the iron that cannot be included in olivine, pyroxene, iron oxides and iron sulphates without violating the abundance constraints (mainly the abundance of silicon and magnesium) is assumed to be in metallic form. Of course, this guarantees a precise reproduction of the iron abundance in dust, but it also means that we cannot rightfully claim to have good constraints on it. Nevertheless, a model without metallic iron does not fit the observed depletion pattern, and so the existence of metallic iron appears necessary, and thus confirmed. The abundance of this species is nonetheless significant; the mass fraction must be roughly 25\%, which is similar to the predicted fractions of olivine, pyroxene, and iron oxides (see Fig. \ref{massfractions} and Fig. \ref{massfractions_mod}). This result is expected and confirms that the SGM cannot be correct.

Even if oxygen, sulfur, and iron are not well-constrained, they still add valuable information about the composition of cosmic dust. An example of this is the large fraction of metallic iron, which is indeed uncertain but definitely not insignificant. If we take the abundances in dust found in Paper I at face value, even a model which consists of only magnesium-free silicates [fayalite (FeSiO$_2$), ferrosilite (FeSiO$_3$)] will not account for all the depleted iron. Assuming, for instance, that we put as much as possible of the oxygen in fayalite and the remainder in w\"ustite (FeO), then the iron depletion can be explained. The depletion of magnesium however will in this case be difficult, if not impossible, to explain with other magnesium-bearing species. More precisely, if the remaining oxygen goes into MgO or some other magnesium-bearing species, this would then explain the depleted magnesium for [Zn/Fe] = 1.2, 1.6, but not for lower [Zn/Fe]. The same phenomenon will occur even with a more realistic composition model without metallic iron, which is why the Monte Carlo simulations favor compositions with a significant, albeit uncertain, fraction of iron  not in silicates and iron oxides.

\subsection{Degeneracies}
A dust-composition model based on a measured depletion pattern is never completely free of degeneracies, because a given depletion pattern can sometimes be explained with vastly different dust compositions. This problem can be seen, to some extent, also in this paper and cannot be resolved without additional information, such as extinction curves for example. The fundamental problem is that there can always be more unknowns (dust species) than there are equations (for considered elements) because of, for example, stoichiometry parameters such as $x_{\rm py}$ and $x_{\rm ol}$, as well as the fact that there are several ways in which two (or more) elements can form molecules.  Another type of degeneracy (mathematical singularity) occurs in our model  when $x_{\rm py} = 2x_{\rm ol}$. This means that the iron-content proportions in silicates sufficiently close to the line $x_{\rm py} = 2x_{\rm ol}$ in the $x_{\rm py}$ -- $x_{\rm ol}$ plane cannot be considered. Therefore, we cannot argue that we have found a single specific composition that explains the depletion patterns, although we may have gained a somewhat better understanding of the major components of noncarbonaceous cosmic dust.

\section{Summary and conclusions}
We use observed depletions of oxygen, sulfur, silicon, magnesium, and iron due to the presence of dust in the ISM to infer the composition of the dust content of the Galaxy and DLAs, down to low metallicity and intermediate redshift. The dust composition is derived computationally by finding the statistically expected elemental abundances in dust assuming a set of a key dust species, with the iron content in silicates and the abundance of metallic iron grains as free parameters. This has been achieved by stochastic (Monte Carlo) simulation of different composition types including both iron-poor and iron-rich silicates, iron oxides, sulfides, and metallic iron.

Despite the limitations of the model, we can robustly conclude the following.
\begin{itemize}
\item There is no single dominant silicate type in the overall dust-mass compositions of DLAs; the interstellar silicates are likely a mixture of both iron-poor and iron-rich olivine and pyroxene. 
\item The overall composition of carbon-free dust is not likely to vary dramatically with the overall dust content and metallicity.
\item A significant fraction of dust (roughly 25\%) is in the form of iron oxides, and a similar fraction is in the form of  metallic iron, perhaps in the form of inclusions in silicates. Thus, iron and iron oxides make up a significant part of the mass of carbon-free dust. The dominant iron oxide is unknown, but w\"ustite (FeO) is a likely candidate since it can form relatively easy.
\end{itemize}
The conclusions above will be further investigated in our forthcoming paper on the expected extinction curves in different environments given different dust compositions (Paper III).

\begin{acknowledgements} 
The reviewer is thanked for his/her constructive criticism which helped improve the paper.
This research was partly financed by the Swedish Research Council, grant no. 2015-04505. The authors also wish to thank Darach Watson for his insightful comments on a preliminary version of the manuscript.
\end{acknowledgements}

\bibliographystyle{aa}
\bibliography{refs_dust.bib}{}

\appendix
\section{Dependence of silicate fractions on the iron content}
This section contains plots of the mass fractions of olivine and pyroxene as functions of the magnesium/iron parameters $x_{\rm ol}$ and $x_{\rm py}$, which are referred to in the main text. We show only the case of an unmodified composition of type A1, since the other composition types lead to very similar (almost identical) results.

The dashed line indicates $x_{\rm ol} = 2\,x_{\rm py}$, where the chemistry matrix may become singular and no meaningful solutions to the Eq. (\ref{matrix}) can be generated. We have carefully designed the numerical algorithm to avoid the singularities. However, near this line numerical artefacts may still occur, because even with this ``sidestepping'' by the algoritm, that is, an introduction of a small shift in the $x_{\rm ol}$ and/or $x_{\rm py}$ values whenever $x_{\rm py} = 2\,x_{\rm ol}$, the LU decomposition used in the matrix inversion routine may sometimes fail in the vicinity of this critical line as a consequence of the fact that ${\rm det}(\mathbf{ X})\approx 0$.  

      \begin{figure*}
    \resizebox{\hsize}{!}{
  \includegraphics{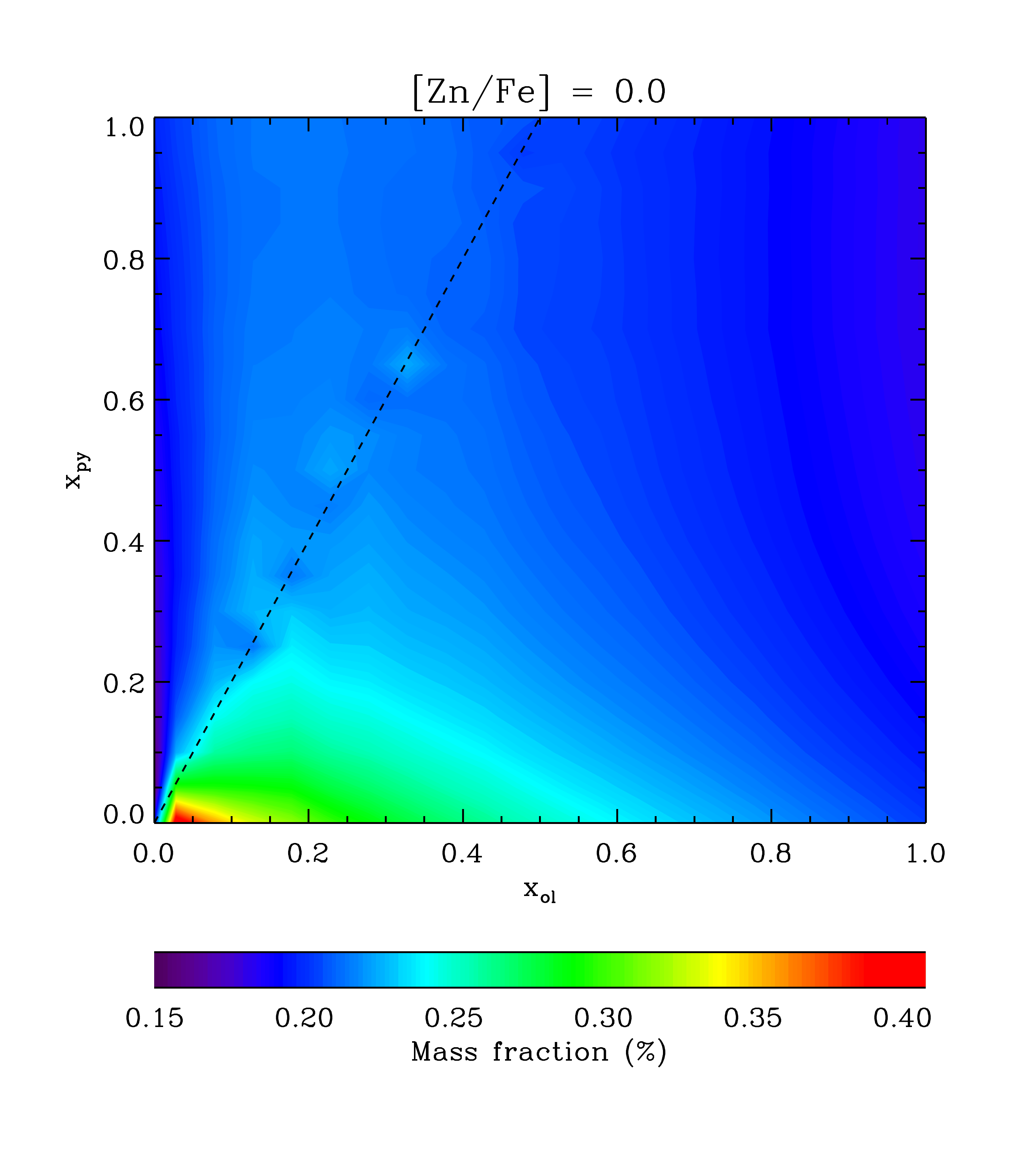}
  \includegraphics{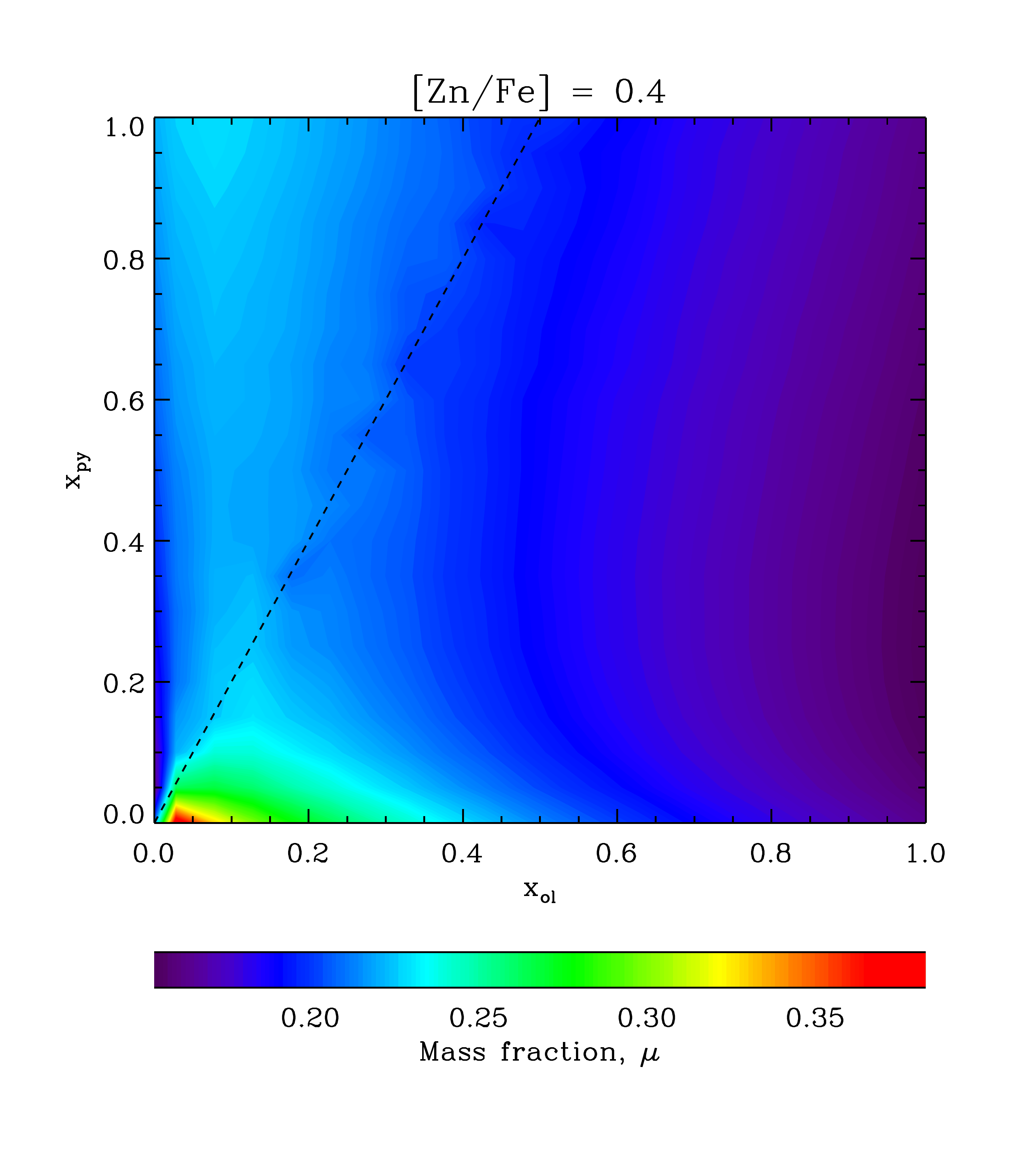}
  \includegraphics{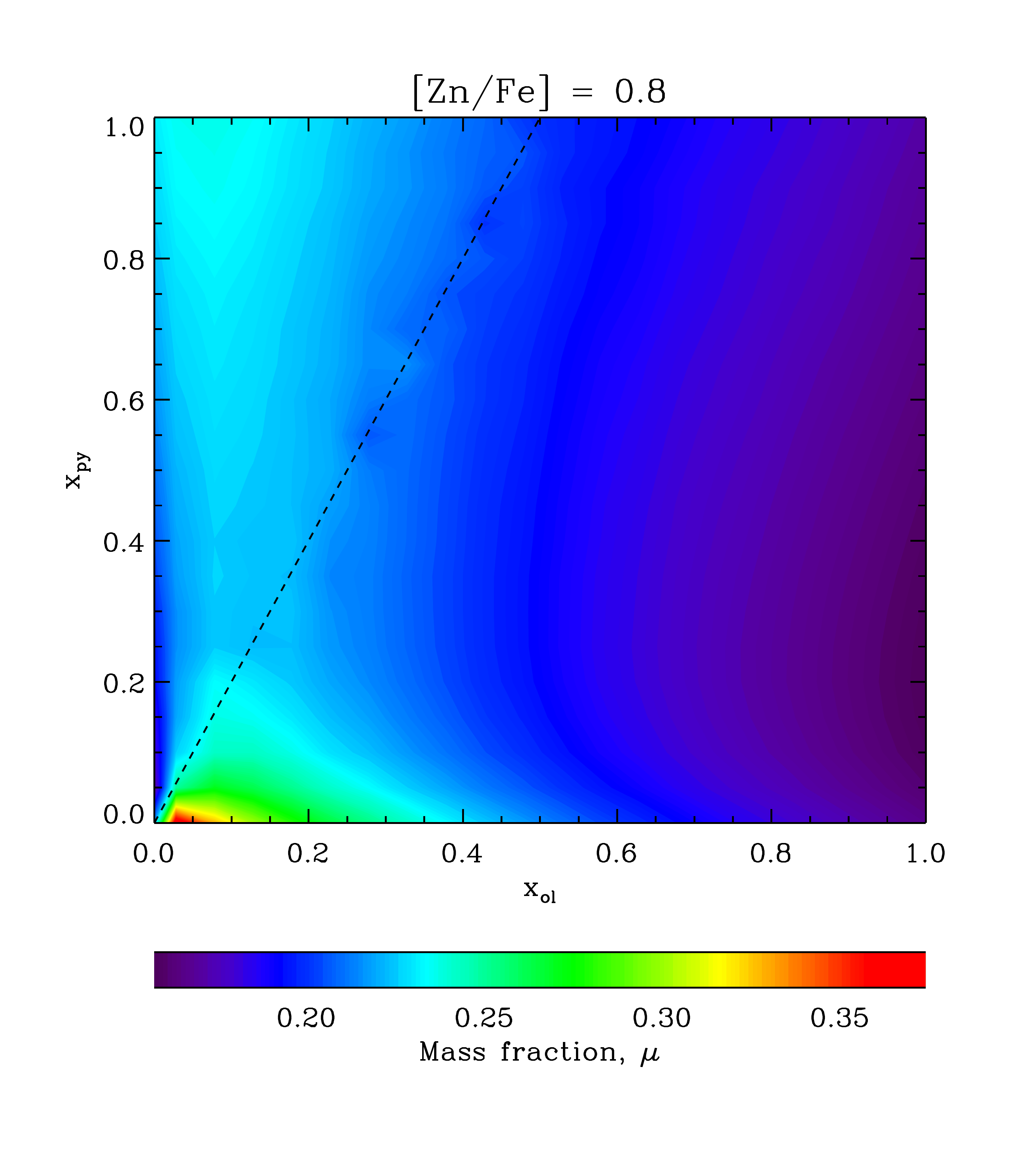}}
  \resizebox{\hsize}{!}{
  \includegraphics{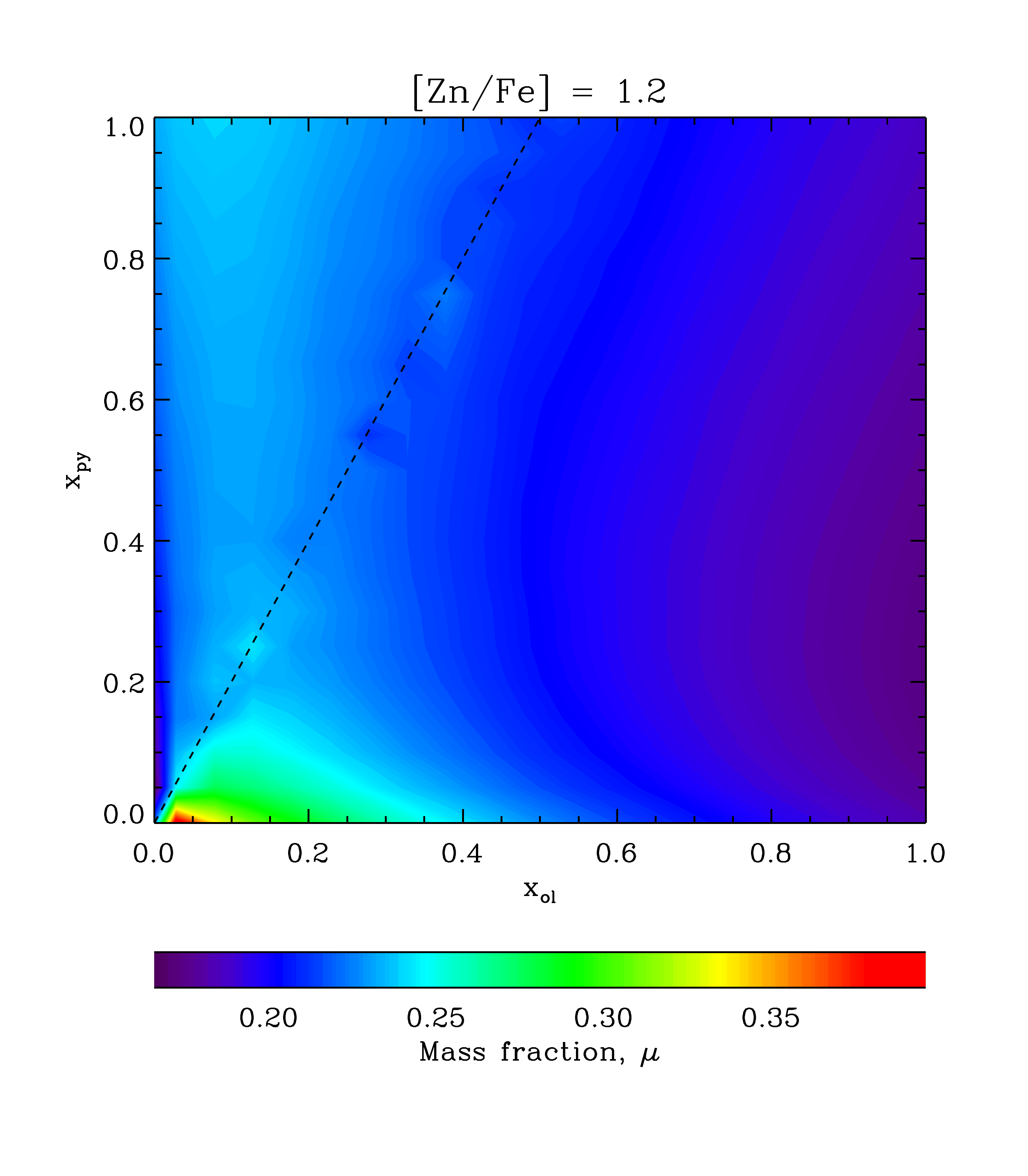}
  \includegraphics{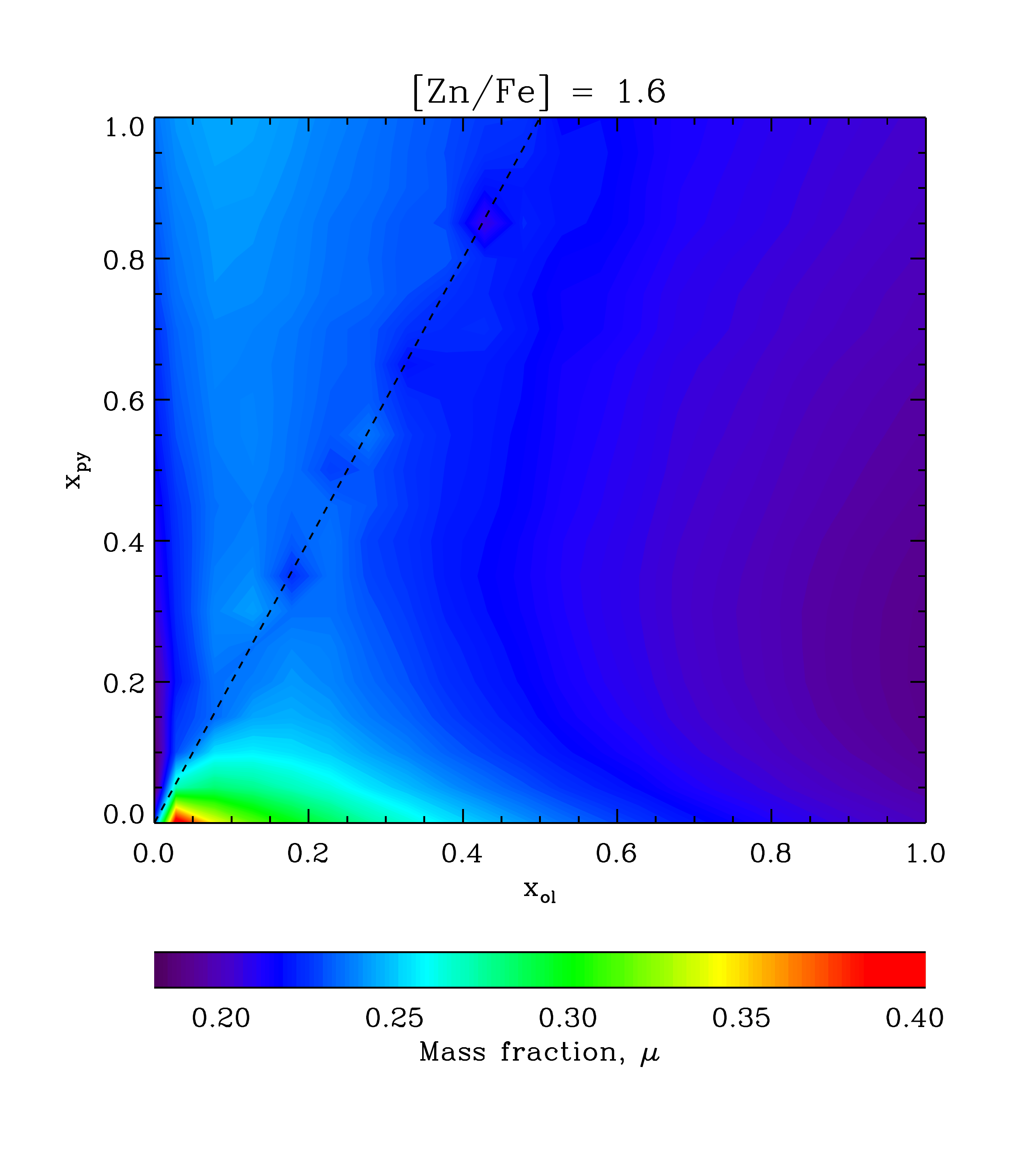}
   \includegraphics{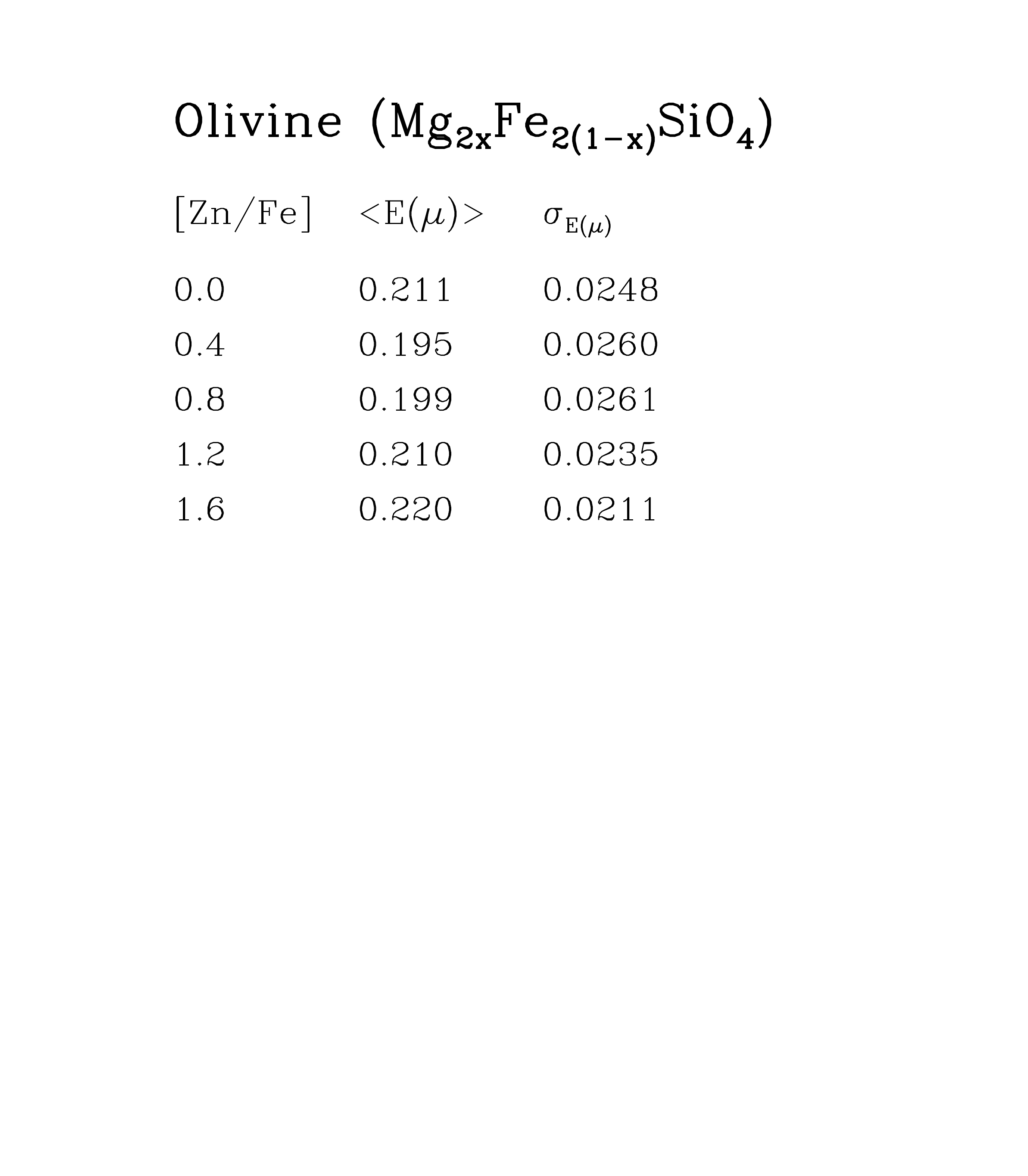}
  }
  \caption{Mass fraction of olivine as a function of the magnesium/iron parameters $x_{\rm ol}$ and $x_{\rm py}$ as predicted using an unmodified composition of type A. The dashed line indicates $x_{\rm ol} = 2\,x_{\rm py}$, where the chemistry matrix may become singular. The numerical algorithm is avoiding the singularities, but near this line numerical artefacts may occur. \label{modelA_ol}
   }
  \end{figure*}
  
      \begin{figure*}
    \resizebox{\hsize}{!}{
  \includegraphics{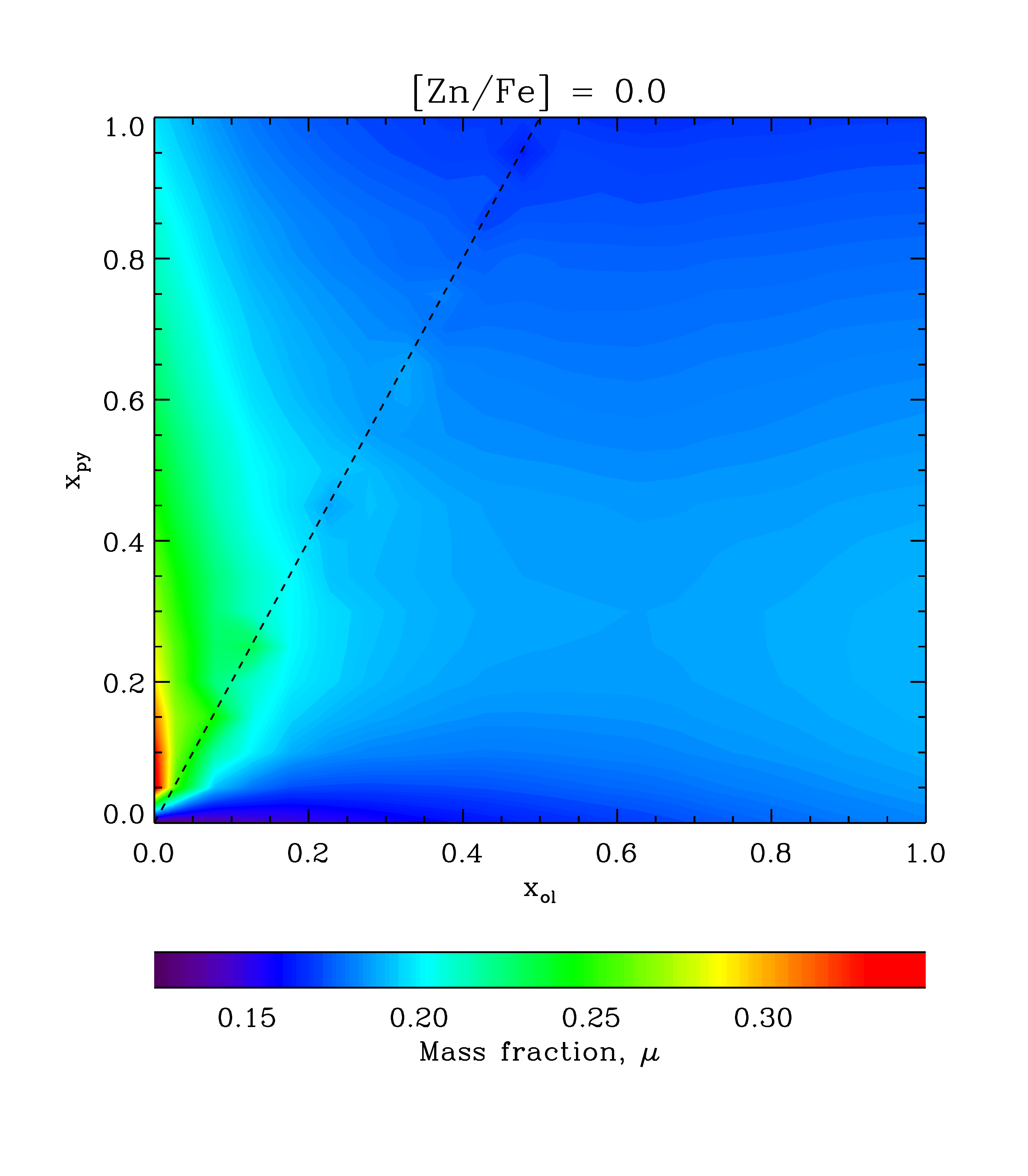}
  \includegraphics{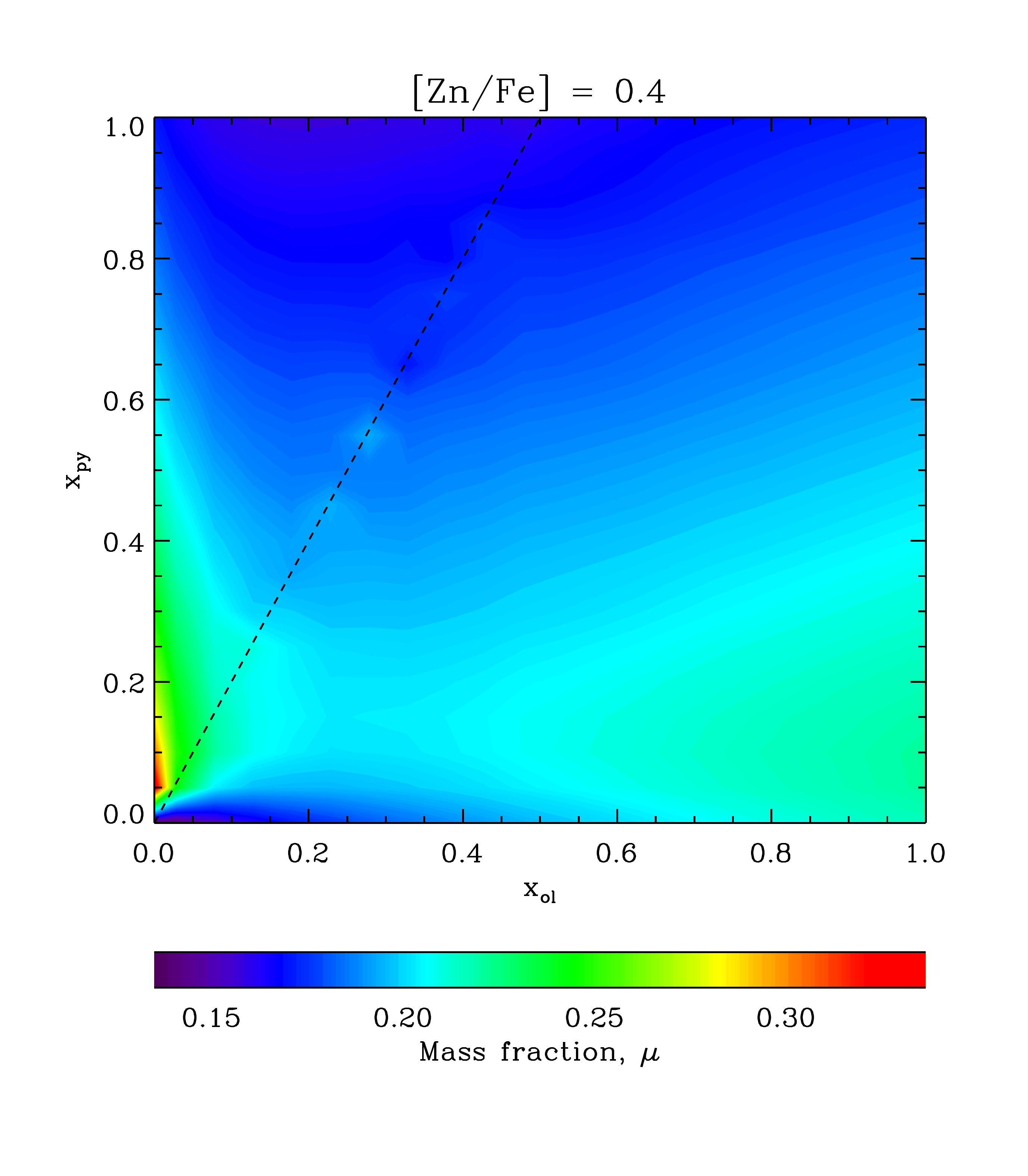}
  \includegraphics{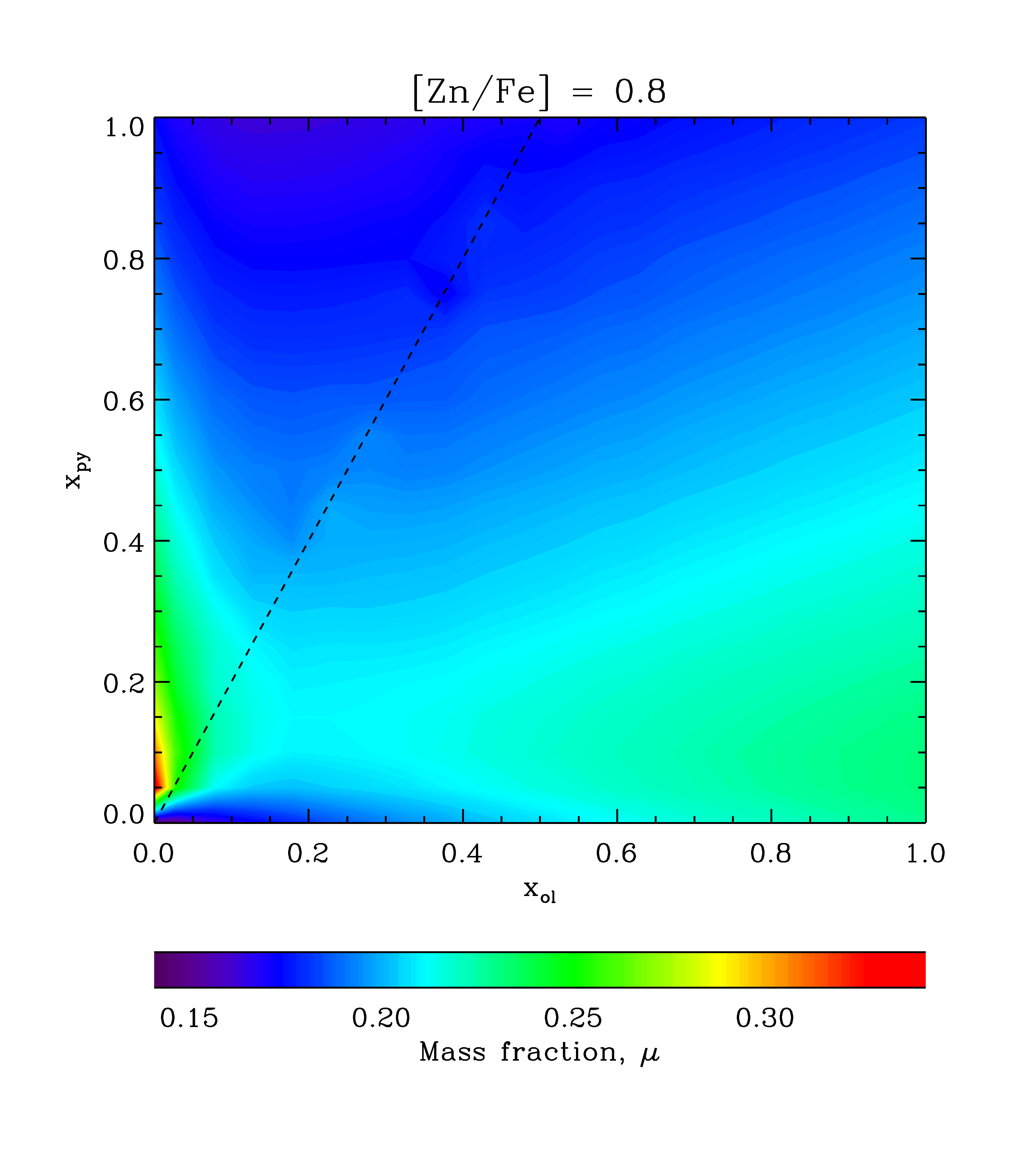}}
  \resizebox{\hsize}{!}{
  \includegraphics{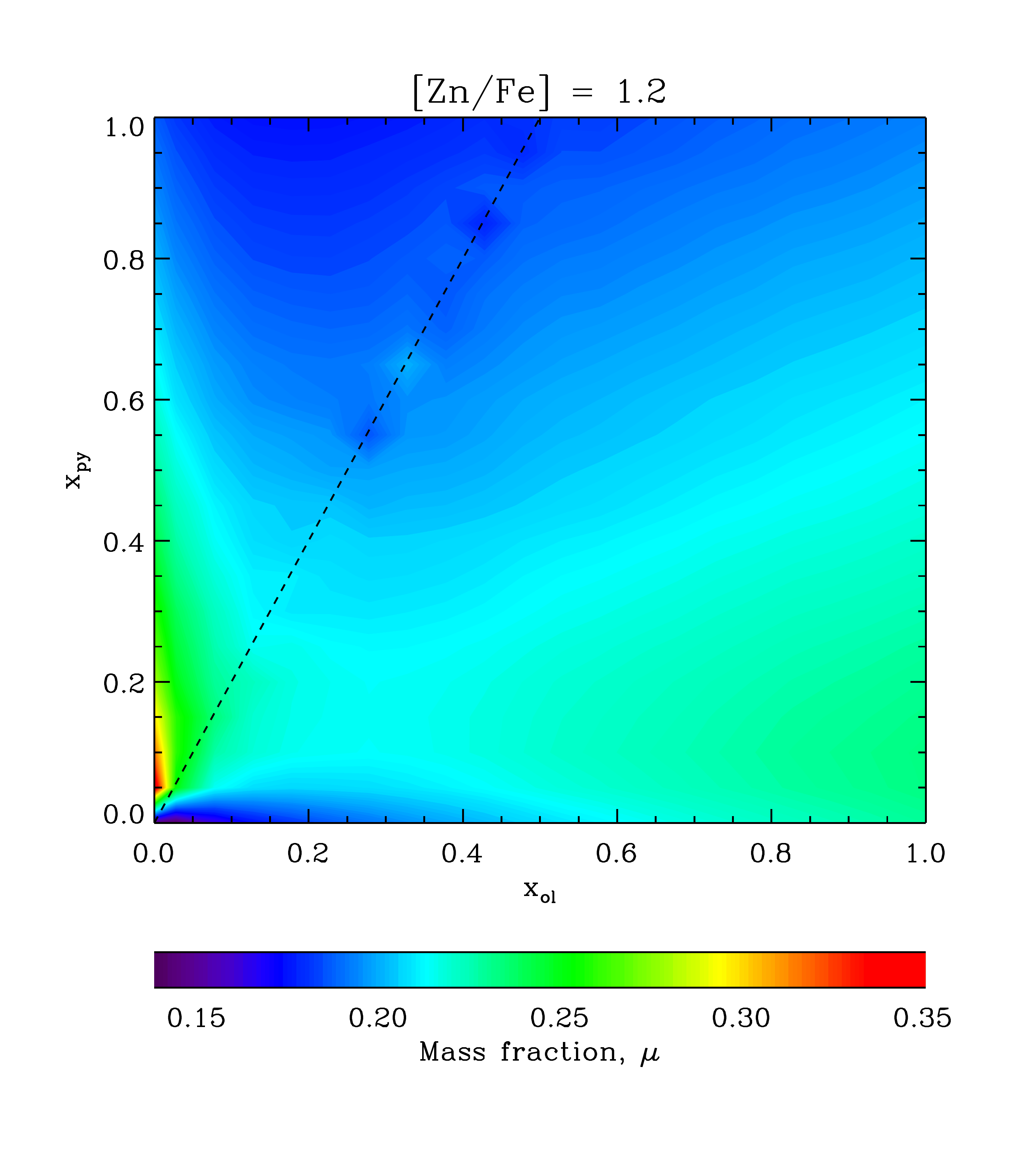}
  \includegraphics{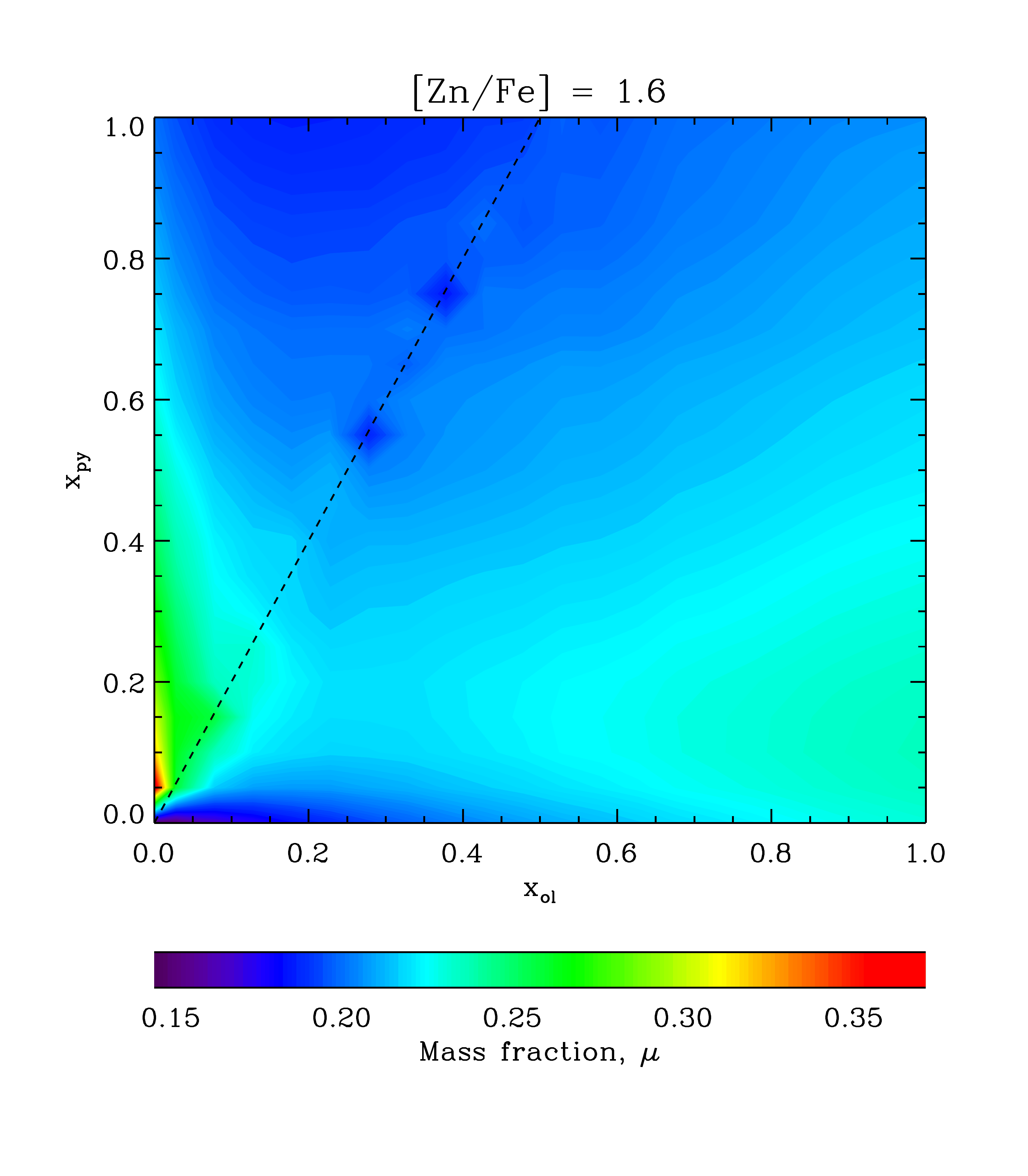}
   \includegraphics{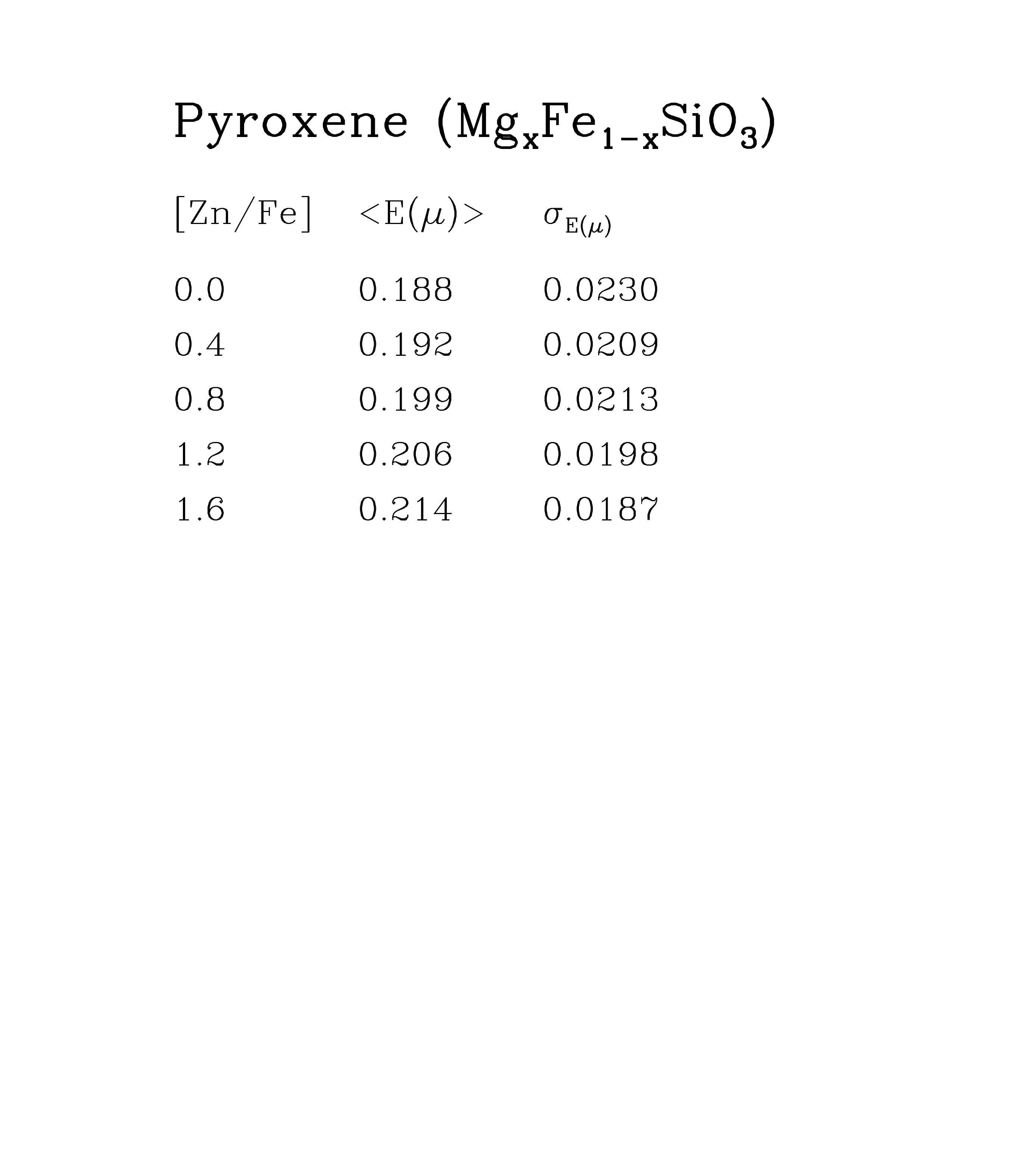}
  }
  \caption{Mass fraction of pyroxene as a function of the magnesium/iron parameters $x_{\rm ol}$ and $x_{\rm py}$ as predicted using an unmodified composition of type A.   \label{modelA_py}
   }
  \end{figure*}

\end{document}